\documentclass[hidelinks,10pt]{article}
\usepackage[utf8]{inputenc}
\usepackage[english]{babel}    
\usepackage{graphicx}
\usepackage[a4paper,lmargin={2cm},rmargin={2cm},tmargin={2.5cm},bmargin={2.5cm}]{geometry}
\usepackage{caption}
\usepackage{subcaption}
\usepackage[onehalfspacing]{setspace}
\usepackage{latexsym}
\usepackage{relsize}
\usepackage{rotating}
\usepackage{ragged2e}
\usepackage[toc,page]{appendix} 
\usepackage{amsmath}
\usepackage{mathtools}
\usepackage{enumitem}
\usepackage{lipsum}
\usepackage{color}
\usepackage{url}
\usepackage{amssymb,amsfonts}
\usepackage{bbold}
\usepackage{amsthm}
\everymath{\displaystyle}
\usepackage{float}

\usepackage[flushleft]{threeparttable}

\newenvironment{fignote}{\begin{quote}\footnotesize}{\end{quote}}

\usepackage{booktabs}
\usepackage[svgnames]{xcolor}
\usepackage[pdftex,bookmarks,colorlinks]{hyperref}
\hypersetup{
    colorlinks,
    citecolor=RoyalBlue, 
    linkcolor=ForestGreen,
    urlcolor=NavyBlue}
    
\usepackage{appendix}
\usepackage{natbib} 
\usepackage[export]{adjustbox}
\usepackage{pdflscape}		%	Make a landscape page
\usepackage{multirow}       %   Let symmetric-entry span over multiple rows
\usepackage{colortbl}
\definecolor{siena}{rgb}{0.91,0.45,0.32} 

\title{\singlespacing Zombie-Lending in the United States \\
\Large{Prevalence \textit{versus} Relevance}}

\author{Maximilian Göbel \thanks {Corresponding author. E-mail: \href{maximilian.goebel@phd.iseg.ulisboa.pt}{maximilian.goebel@phd.iseg.ulisboa.pt}. This work was supported by the FCT (\textit{Funda\c{c}\~{a}o para a Ci\^{e}ncia e a Tecnologia}) [grant number UIDB/05069/2020].
REM – Research in Economics and Mathematics, UECE – Research Unit on Complexity and Economics.} \\ ISEG - Universidade de Lisboa, UECE, REM
	\and
	 Nuno Tavares \thanks {E-mail: \href{nuno.tavares@phd.iseg.ulisboa.pt}{nuno.tavares@phd.iseg.ulisboa.pt}} \\ ISEG - Universidade de Lisboa}

\date{\vspace{0.75cm} \small First Draft: September 19, 2021 \\ This Draft: \today \\ \vspace{0.5cm}
}
\begin{document}
	\begin{titlepage}
		\maketitle
		\begin{abstract}
			\singlespacing   
			
Extraordinary fiscal and monetary interventions in response to the COVID-19 pandemic have revived concerns about zombie prevalence in advanced economies. Within a sample of publicly listed U.S. companies, we find zombie prevalence and zombie-lending not to be a widespread phenomenon per se. Nevertheless, our results reveal negative spillovers of zombie-lending on productivity, capital-growth, and employment-growth of non-zombies as well as on overall business dynamism. It is predominantly the class of healthy small- and medium-sized companies that is sensitive to zombie-lending activities, with financial constraints further amplifying these effects.

			\vspace{0.2cm}
			
			\vskip 10pt
			\noindent\textbf{Acknowledgements:} For helpful comments we thank, without implicating, Viral V. Acharya, Philippe Goulet Coulombe, Panagiotis Avramidis, Stefano Colonnello, Tommaso Oliviero, Petru Crudu and the participants of the Finance Workshop 2022 at Ca' Foscari University of Venice.
		\end{abstract}

	\end{titlepage}
\newpage

\pagenumbering{arabic}

\section{Introduction}

The COVID-19 pandemic constitutes the single most disruptive event the world economy has experienced since the second world war. The tremendous health and safety challenges posed by the pandemic, forced governments to undertake unprecedented action to contain the spread of the disease. Many countries adopted strict lockdown policies that constrained firms in nonessential sectors to shut down completely. Policy responses designed to keep businesses afloat included public support on firms’ liquidity, such as wage bills and tax relief schemes, moratoriums on credit installment payments, and credit guarantees. Although there is a large consensus about the need for measures that help flattening the curve of corporate insolvency of otherwise viable firms, there is increasing concern that such measures combined with ill-designed screening schemes might have allowed resources to flow into non-viable firms commonly known as zombies \cite{BoddinDAcuntoWeber2020}.\footnote{\cite{HoshiKawaguchiUeda2022} find evidence for the case of Japan, where poorly performing firms made frequent use of government lending programs. Yet, these firms do not necessarily fall under their zombie definition. } 

\noindent The term``zombie'' was first introduced by \cite{CaballeroHoshiKashyap2008} in their analysis of Japan's \textit{"lost decade"} of the 1990s and is often used to describe firms that are consistently unable to generate enough profits to meet their debt-servicing expenses. The literature on the topic has been consistent in identifying zombies and zombie-lending as drivers of productivity slowdown, either by stifling overall productivity growth or by intensifying misallocation of resources in the economy \citep{CaballeroHoshiKashyap2008, McGowanAndrewsMillot2018}. The literature also points to the fact that zombie-enduced congestion contributes to declining profits of healthy firms (non-zombies), discouraging investment and entry of new firms (newbies) (see for instance \cite{hallak2018fear} for the European case).  The rise of zombies is often associated with weaknesses stemming from the financial sector \citep{AndrewsPetroulakis2019}. Weaker banks, overflowed with doubtful assets sunken into zombies, may have strong incentives to engage in debt evergreening practices, allowing them to roll over loans instead of writing them off, thus exacerbating resource misallocation \citep{peek2005unnatural,storz2017we}.

\noindent Extraordinary monetary stimulus and lingering crisis-driven support for small- and medium-sized enterprises (SME) was the European response to the Global Financial Crisis (GFC). Combined with institutional factors, such as poor insolvency regimes, these policies have raised concerns that non-viable firms are kept alive artificially \citep{McGowanAndrews2018}. 
%In Europe, extraordinary monetary stimulus, lingering crisis-driven support for small- and medium-sized enterprises (SME) in the aftermath of the Global Financial Crisis (GFC) as well as institutional factors, such as poor insolvency regimes, have triggered concerns that non-viable firms are kept alive artificially \citep{McGowanAndrews2018}.
Recent events, prompted by the pandemic, have further stimulated the debate about whether these actions, although justifiable, may fuel the zombie phenomenon, as the need to act fast and decisively reduces incentives and increases the difficulty to accurately screen the creditworthiness of borrowers \citep{LaevenSchepensSchnabel2020}. 
The U.S. economy is no exemption. The unprecedented fiscal and monetary policy support in the wake of the COVID-19 pandemic has triggered similar fears of firm zombification\footnote{See for instance: \href{https://www.washingtonpost.com/business/2020/06/23/economy-debt-coronavirus-zombie-firms/}{\textit{"Here’s one more economic problem the government’s response to the virus has unleashed: Zombie firms"} \textcolor{black}{(Washington Post, June 23, 2020)}; \href{https://www.ft.com/content/9b304e20-49cf-4fba-81a0-4d06f930d7a1}{\textit{"Pandemic debt binge creates new generation of ‘zombie’ companies"} \textcolor{black}{(Financial Times, September 13, 2020)}}}}.  \cite{FavaraCameliaPOrive2022} find that zombie prevalence is not widespread among private and publicly listed firms in the U.S., which leads them to infer that zombie firms only have a minor impact on aggregate productivity and credit misallocation in the U.S. economy. %However, they also note that it is too early to dismiss concerns that the current economic conditions may be breeding ground for new zombie firms. 
While \cite{FavaraCameliaPOrive2022} set the primary focus on banks' engagement in zombie-lending, our objective is to understand the potential spillover effects of granting credit to zombies on healthy firms.

\noindent In other parts of the world, the literature has already explored the zombie phenomenon much more thoroughly. The evidence is rather unambiguous. Several studies on Japanese and European firms have found zombification to constrain the flourishing of non-zombies, leading to diminished growth in productivity, investment, and employment \citep{CaballeroHoshiKashyap2008, McGowanAndrewsMillot2018, AndrewsPetroulakis2019, AcharyaEtAl2019}. Others go even further by noting that zombie prevalence interferes with an efficient transmission of monetary policy \citep{AcharyaEtAl2020}. 
Despite these findings, little evidence has been brought to bear about the influence of zombies and, in particular zombie-lending, on economic outcomes in the U.S.\footnote{\cite{BanerjeeHofmann2020} include the United States in their analysis of zombification in 14 advanced economies, but focus more on the life cycle of zombies across several countries than on the micro- and macroeconomic consequences of zombification -- and in particular lending to zombies.} In this paper, we intend to address this gap by analysing potential spillovers of zombification -- in particular zombie-lending -- on productivity, investment, employment-growth, and business entry dynamics. 
We merge firm-level data of publicly listed firms, sourced from Compustat and Standard \& Poor's Capital IQ for the period of 2002-2020, to identify non-viable firms and their debt structure. This data set is in so far appealing as it allows us to not only distinguish between debt, intermediated by banks, and credit, taken up via the bond market, but to also discriminate between short- and long-term funding.

%Having information on both origins of funding constitutes a unique feature, allowing us to delve deeper into different layers of interaction between the zombie status, sources of financing, and firms' outcomes. 

\noindent Our contribution to the literature on zombification is threefold. Second to \cite{FavaraCameliaPOrive2022} we are among the first to provide an overview of the degree of zombie prevalence among publicly listed U.S. companies. %We identify zombie firms using common definitions proposed in the related literature. Specifically, we 
We construct two alternative measures of zombie identification, a broader version that operationalizes the concept based on \cite{McGowanAndrewsMillot2018}, and a narrower version that includes expectations about future profitability \citep{banerjee2018rise}. Both definitions reveal a steady increase in the share of zombie firms in the U.S. dating back to the mid-'90s. After peaking during the GFC, the share of zombies started to dwindle from 2010 onwards, although still far from pre-crisis levels. Yet, zombie prevalence does not seem to be a defining feature in the U.S., particularly if we use more stringent and demanding criteria to access the zombie status of firms. no
Our second contribution differentiates us from \cite{FavaraCameliaPOrive2022}, who pose more emphasis on analyzing banks' zombie-lending engagements. We, however, make an attempt at answering the question whether the lack of prevalence translates into the irrelevance of zombies when it comes to economic outcomes. Our results suggest it does not. We find negative spillover effects of zombie-lending on the performance of healthy companies, emerging both from public debt markets as well as from the traditional bank-lending channel. The impact is most pronounced for small- and medium-sized firms, corroborating the findings in \cite{BanerjeeHofmann2020}. Bank-dependency and lack of access to the bond market further amplify the negative effects, suggesting financial constraints to be a catalyst for the negative spillovers of zombie-lending. %Furthermore, a differentiation between short- and long-term lending is worthwhile for identifying the different transmission channels. %as different performance measures seem to be maturity sensitive. 
Lastly, our findings add to the literature on business dynamism. The results suggest that inflows of zombie-credit from the banking sector interfere with entry and exit dynamics at the two-digit NAICS industry level.

\noindent The remainder of this paper is organized as follows. Section \ref{sec:literature} discusses the related literature. Section \ref{sec:Data} describes the data used, while Section \ref{sec:ZombieDef} discusses the identification of zombie firms. Section \ref{sec:Zombie_Evidence} outlines our empirical approach and our chieftain results. Section \ref{sec:concl} concludes.

%Our empirical approach focuses primarily on the effect of \zombie prevalence and \textit{zombie lending} on firm-level productivity, investment, and employment growth of U.S. firms during this period. (report a summary of main findings without much detail). We close by examining the impact of \textit{zombie lending} on one of the most relevant transmission channels, i.e. business dynamism (report a summary of main findings without much detail).

\section{Literature Review}\label{sec:literature}

\noindent The debate about zombie firms can be traced back to  \cite{CaballeroHoshiKashyap2008} who first introduced the concept in their study on Japan's \textit{"lost decade"}. They argue that the prevalence of this type of firm depressed market prices and increased wages relative to productivity, preventing healthy firms from flourishing by stifling the creative destruction process. Building upon this work, a recent series of papers have tracked the zombie phenomenon among OECD countries. For instance, 
\cite{BanerjeeHofmann2020} describe the life cycle of zombies across 14 advanced economies -- which also comprises the United States. In their sample of listed companies, zombies make up about 6\%-7\% of total assets, capital and
debt and is found to be a particularly widespread phenomenon among SMEs. 
Although a large proportion of zombies can eventually recover, they remain weaker and more fragile than their peers, which have never been classified as non-viable. 
Despite the aggregate numbers appearing to be minor, the authors warn against the perception that zombie prevalence won't have a saying in the future trajectory of the economy.
\cite{McGowanAndrewsMillot2018} show that the prevalence of resources sunk in zombie firms have risen since the mid-2000s and that the increasing survival of these low productivity firms at the margins of exit congests markets and constrains the growth of more productive firms. Furthermore, they show that a higher share of industry capital sunk in zombie firms lowers investment and employment growth of non-zombies. Similarly, \cite{AcharyaEtAl2020} conclude that zombie-lending results in a misallocation of capital that culminates in lower product prices, productivity, investment, and value added.
%We add to this literature by, firstly, providing a general description of zombie prevalence in the United States under various classification schemes, and secondly, by discussing the implications of zombie-lending -- both from the banking sector and via capital markets -- for the performance of non-zombies.
%Our analysis differs in that we focus exclusively on the United States. Rather than describing the life cycle and detailed characteristics of zombies, we study the micro- and macroeconomic consequences of zombification and in particular of lending credit to zombies using data on the debt structure of individual firms. Thereby, we distinguish not only between credit granted by banks or capital markets, but also between the maturity of a particular debt instrument.

\noindent A defining feature of the literature on zombies relates to the problem of identification using firm-level data. Whereas, in economic terms, zombies can be defined as non-viable firms that would exit the market in the absence of frictions, the operationalization of the concept is not straightforward. For instance, \cite{CaballeroHoshiKashyap2008} consider a firm to be a zombie if it has continued access to financial support from their creditors, despite their poor performance in terms of profitability. To make this definition operational, the authors compare the interest rate paid by the firm with a benchmark rate applied to high-quality borrowers. Those firms that present a negative interest gap receive subsidized credit and are thus classified as zombies. This method, although feasible, is very demanding in terms of data that is rarely available. A similar approach to zombie-firm classification can be found in \cite{AcharyaEtAl2019, AcharyaEtAl2020}.
Alternatively, a definition that is often used as a benchmark definition \citep{McGowanAndrewsMillot2018,banerjee2018rise,AndrewsPetroulakis2019}, only requires knowledge of a firm's age and its interest coverage ratio (ICR). This zombie metric is less demanding in terms of data availability, though \cite{AndrewsPetroulakis2019} still remark poor coverage of a firm's interest expenses in certain data sets.
To get around a potentially sparsely populated ICR, the literature also classifies a firm as being a zombie, if -- over three consecutive years -- the firm reports a low debt-service capacity, and either negative return on assets (ROA) or negative capital-growth \citep{storz2017we,AndrewsPetroulakis2019}.
In Section \ref{sec:ZombieDef}, we discuss the implication of only using a firm's age and its ICR as the defining features in the U.S. case and argue in favor of a slight modification.

\noindent Another strand of the literature, which our underlying data does not allow us to follow up on, addresses the motivation to engage in zombie-lending from a lender's perspective.
On the one hand, institutional bottlenecks such as poor insolvency regimes are found to be important enablers of zombification \citep{mcgowan2017insolvency}, constituting a mix which aggravates the macroeconomic effects of corporate credit booms going bust \citep{JordaKornejewSchularickTaylor2020}. On the other hand, an increasing body of literature links the frailties of the financial sector to the rise of zombies. Early work by \cite{peek2005unnatural} already found that weaker firms were more likely to receive additional credit, because troubled Japanese banks were incentivized to allocate funds to severely impaired borrowers in order to avoid the realization of losses on their own balance sheets. Observing lender-borrower relationships in Italy, \cite{schivardi2017credit} find that undercapitalized banks were less likely to cut credit to zombie firms, whereas for the broader European case, \cite{AndrewsPetroulakis2019} show that non-viable firms are more likely to be connected to weak banks, suggesting that zombie prevalence in Europe may at least partly stem from bank forbearance. In spite of growing regulatory pressure, there is evidence that zombie-lending remains widespread, even in developed countries
\citep{bonfim2020site}. 

\noindent This stands in contrast to the findings in \cite{FavaraCameliaPOrive2022} -- the most closely related study in terms of examining zombie prevalence in the United States. Their finding of a low degree of zombification in the U.S. economy extends to zombie lending not being a widespread phenomenon among private and publicly listed companies. They find banks to tighten their lending standards or even reduce their exposure to firms, which slide into zombie-like status. Furthermore, they don't find any support for undercapitalized banks engaging more forcefully in zombie-lending than other financial institutions, which contrasts with the findings from the European case \citep{schivardi2017credit,AndrewsPetroulakis2019} .
This seminal contribution on banks' zombie-lending activities in the United States provides valuable insights into the functioning of the U.S. banking market, and how it differs from its European counterpart.

%Our findings corroborate the former, but stand in contrast to the latter. 
We differ from \cite{FavaraCameliaPOrive2022} by analysing the interplay between zombie-lending and the performance of non-zombie firms. We find statistically significant negative spill-overs on productivity, investment, and employment growth. Nonlinearities are a crucial part of the equation: while large companies seem to be unaffected, it is SMEs -- and subgroups thereof -- that are the ones to carry the entire burden of zombie lending.
These results go hand in hand with findings in \cite{Karakaplan2021} and \cite{ChodorowReichEtAl2021}, in that they corroborate the existence of large heterogeneity and nonlinearities in the cross-section of U.S. companies and emphasize the implications for small- and medium-sized enterprises.

\noindent Lastly, we add to the literature on business dynamism by examining the link between zombie-lending and the rate of entrance at the industry-level. Studying zombie firms and specifically \textit{"zombie-credit"} in Europe, \cite{AcharyaEtAl2020} find that increased lending to zombies hampers the cleansing effect in the economy, i.e. the replacement of non-viable firms by new entrants. We find similar effects for the U.S., with the main spillovers emerging from bank-related zombie-lending. We also find support for a working hypothesis -- also outlined in \cite{AcharyaEtAl2020} -- explaining one channel through which zombie-lending compromises market entry.

\noindent Before turning to the empirical analysis, the next two sections provide further details on the characteristics of the data and zombie prevalence among publicly listed companies in the United States.

%Other frameworks require a firm to report negative profits or negative value added \cite{McGowanAndrewsMillot2018}.

\section{Data}\label{sec:Data}
Our empirical analysis is based on annual firm-level observations of public U.S. companies. For information on companies' balance sheets we resort to Compustat's annual files, while Standard \& Poor's Capital-IQ (CapIQ) database is our source for companies' debt instruments. The benefit of these two databases is the one-to-one mapping between the debt instruments listed in CapIQ and a company's fundamentals in Compustat's annual files. The drawback of CapIQ is its poor coverage prior to 2002, which shrinks our effective sample size to 19 years between 2002-2020.\footnote{Another source of information about a company's lending-relationships is \textit{DealScan}, which reports syndicated loan arrangements with borrower and lender identification. However, a direct mapping between DealScan and Compustat is not given without a little detour. Further, syndicated loan agreements usually come with large face-values. As further analyses will show, zombie-firms are rather small compared to their peers in the Compustat sample. Thus, based on further examinations, zombies (even when following along our most generous zombie-definition) are only sparsely represented in the DealScan data set.} 
We consider all firms in the Compustat database with an identifier $fic = USA$. Furthermore, we exclude sectors with NAICS codes 11, 22, 52, 55, 81, and 92. We deflate financial variables by the industry-specific producer price index derived from \textit{KLEMS}\footnote{Source: \textcolor{blue}{\href{https://www.bea.gov/data/special-topics/integrated-industry-level-production-account-klems}{bea.gov}}} with base year 2002.\footnote{We do neither winsorize the data nor omit any percentiles in our empirical analysis.}

\subsection{Company Financials} 
Our goal is to evaluate the impact of zombie-lending on non-zombie firms. To measure the relative performance of non-zombies, we focus in particular on productivity, capital- and employment-growth. Appendix \ref{sec:app_variables} describes the computation of Total Factor Productivity ($TFP$) and Appendix \ref{sec:app_variables} contains a description of the variables used in the empirical part of the paper.

%\noindent Here, we only shortly outline the computation of two of our main variables of interest: labor productivity ($LP_{i,t}$) and total factor productivity ($TFP_{i,t}$). We measure labor productivity of firm $i$ as the amount of value added ($VA_{i,t}$) -- measured in million USD -- produced by each employee. We follow standard procedures in the literature \citep{ImTuez2014,HartmanLustigXiaolan2019,Doerr2020}, and define $VA_{i,t}$ as the sum of operating income before depreciation ($OIBDP$) and labor expenses. Since the directly observable measure of labor expenses in the Compustat database ($XLR$) is only sparsely populated, we impute labor expenses for all missing observations with annual labor costs of the U.S. Manufacturing sector. This information is drawn from the Bureau of Labor Statistics (BLS).\footnote{As it turns out, imputing missing information on labor expenses for \textit{all} sectors with data from the U.S. Manufacturing sector only, tracks the \textit{observed} $XLR$ variable more closely than using sector-specific labor expenses. See Appendix \ref{sec:app_variables}, for a further discussion on this issue.}

%\input{Figures/Productivity_Comparison}

\subsection{Debt Contracts} \label{sec:data_debt}
Standard and Poor's Capital-IQ (CapIQ) database allows us to observe both the extent of bank-related credit intermediation in the form of Bank-/Term-Loans ($BL$) and Revolving Credit Facilities ($RC$), and information on firms' financing operations via capital markets in the form of Bonds and Notes ($BN$). Table \ref{tab:SummStats_CapIQ_maturities} shows the total number of debt contracts differentiated by the type of debt and maturity at the time of origination.
Overall, $BN$ account for about 60\% of all observed debt contracts, making it a vital component when trying to paint a comprehensive picture of firm-level debt financing.
In contrast to other studies on zombie-lending in Europe \citep{McGowanAndrewsMillot2018,AndrewsPetroulakis2019,schivardi2017credit}, our data set allows us to not only differentiate between bank-intermediated debt and other debt securities, but also to distinguish between short- and long-term funding.

\noindent For the empirical part, we pre-process the data along the following steps: first, we compute $BC_{i,t}$ as the amount of bank-credit being granted to firm $i$ in year $t$ as the sum of  Bank-/Term Loans $BL_{i,t}$ and Revolving Credit $RC_{i,t}$. From now on, we will thus only distinguish between bank-credit, $BC_{i,t}$, and credit taken on via public debt markets in the form of Bonds \& Notes $BN_{i,t}$. Throughout the empirical analysis of Section \ref{sec:Zombie_Evidence},
the main (explanatory) variable of interest is the industry-share of new debt being granted to zombies in year $t$.\footnote{In CapIQ, debt obligations are identified via their $componentid$. Existing debt obligations only get assigned a new \textit{componentid} if their contract details have changed. However, the same \textit{componentid} may occur multiple times in a company's financial statements if, for example, a downpayment has been made. Henceforth, we only count the reporting of $BL_{i,t}$, $RC_{i,t}$, or $BN_{i,t}$ in year $t$, if its \textit{componentid} has not appeared in any year before $t$ in firm $i$'s filings. %By following this procedure, we capture the willingness of financial institutions and capital market participants to engage both in new financing commitments or re-negotiations of existing contracts
} Therefore, we define $BC_{s,t}$ and $BN_{s,t}$ as the \textit{share} of newly granted credit to industry $s$ in year $t$ sitting with zombies, as follows:

\[ L_{s,t} = \frac{\sum_{i \in S} L^Z_{i,t}}{\sum_{i \in S} L^Z_{i,t} + L^{NZ}_{i,t}}, \qquad \text{for } L = BC,BN \quad , \]

\noindent where $S$ spans the set of two-digit NAICS industries, $Z$ is the zombie indicator and $NZ$ the non-zombie indicator respectively. To reduce the number of reporting errors in the data, we only include an observation, if its face value does not exceed the borrower's total debt, as reported in the Compustat filings.\footnote{Table \ref{tab:SumStat_CapIQ_AcceptanceRates} in Appendix \ref{app:capIQ_stats} documents the corresponding acceptance rates.} 
%This procedure reduces the number of observations either because of spurious reporting or -- unfortunately -- because of missing data in the Compustat files. Table \ref{tab:SumStat_CapIQ_AcceptanceRates} in Appendix \ref{app:capIQ_stats} documents the corresponding acceptance rates.

\noindent Lastly, Table \ref{tab:SummStats_CapIQ_maturities} shows that the majority of new debt obligations is issued with a maturity of less than 10 years. We therefore differentiate between short-term credit as debt obligations with a maturity of less than one year and long-term credit as those with a maturity of more than one but less than 10 years at the date of origination.

\begin{landscape}
    \begin{table}[h!]
    %\begin{center}
    \centering
    \begin{threeparttable}
    \vspace{-1cm}
    \captionsetup{justification=centering}
    \footnotesize
\caption{\label{tab:SummStats_CapIQ_maturities} Total Number of Debt Obligations and their Maturities -- Full Sample: 2002-2020}

\begin{tabular}{l c ccc c ccc c ccc c ccc} \hline
\addlinespace[3pt]
& & \multicolumn{15}{c}{Zombie-Definition: $Z^{NAR}$} \\ 
\cmidrule(lr){3-17}
& & \multicolumn{3}{c}{Bank/Term Loans} & & \multicolumn{3}{c}{Revolving Credit Facility} & & \multicolumn{3}{c}{Bonds and Notes} & & \multicolumn{3}{c}{Total}  \\ 
\cmidrule(lr){3-5} \cmidrule(lr){7-9} \cmidrule(lr){11-13} \cmidrule(lr){15-17} 
& &All & Zombies & Non-Zombies & & All & Zombies & Non-Zombies & & All & Zombies & Non-Zombies & & All & Zombies & Non-Zombies \\ 
\cmidrule(lr){3-3} \cmidrule(lr){4-4} \cmidrule(lr){5-5} \cmidrule(lr){7-7} \cmidrule(lr){8-8}  \cmidrule(lr){9-9} \cmidrule(lr){11-11} \cmidrule(lr){12-12} \cmidrule(lr){13-13} \cmidrule(lr){15-15} \cmidrule(lr){16-16} \cmidrule(lr){17-17} 

%%%%%%%%%%%%%%%%%%%%%%%%%%%%%%%%%%%%%%%%%%%
1Q  $\leq m \leq$ 4Q  & &
14,174 & 517 & 13,657 & &
17,834 & 607 & 17,227 & &
34,302 & 1,065 & 33,237 & &
67,198 & 2,189 & 65,009 \\ \addlinespace[5pt]
%%%%%%%%%%%%%%%%%%%%%%%%%%%%%%%%%%%%%%%%%%%

%%%%%%%%%%%%%%%%%%%%%%%%%%%%%%%%%%%%%%%%%%%
5Q  $\leq m \leq$ 8Q & &
10,831 & 489 & 10,342 & &
12,876 & 511 & 12,365 & &
28,212 & 827 & 27,385 & &
52,230 & 1,827 & 50,403 \\ \addlinespace[5pt]
%%%%%%%%%%%%%%%%%%%%%%%%%%%%%%%%%%%%%%%%%%%

%%%%%%%%%%%%%%%%%%%%%%%%%%%%%%%%%%%%%%%%%%%
9Q  $\leq m \leq$ 20Q & &
30,425 & 1,069 & 29,356 & &
33,630 & 639 & 32,991 & &
75,140 & 1,857 & 73,283 & &
139,830 & 3,565 & 136,265 \\ \addlinespace[5pt]
%%%%%%%%%%%%%%%%%%%%%%%%%%%%%%%%%%%%%%%%%%%

%%%%%%%%%%%%%%%%%%%%%%%%%%%%%%%%%%%%%%%%%%%
21Q  $\leq m \leq$ 40Q & &
18,284 & 465 & 17,819 & &
17,673 & 185 & 17,488  & &
75,449 & 967 & 74,482 & &
111,717 & 1,617 & 110,100 \\ \addlinespace[5pt]
%%%%%%%%%%%%%%%%%%%%%%%%%%%%%%%%%%%%%%%%%%%

%%%%%%%%%%%%%%%%%%%%%%%%%%%%%%%%%%%%%%%%%%%
41Q  $\leq m \leq$ 100Q & &
4,379 & 181 & 4,198 & &
3,602 & 42 & 3,560 & &
35,178 & 411 & 34,767 & &
43,199 & 634 & 42,565 \\ \addlinespace[5pt]
%%%%%%%%%%%%%%%%%%%%%%%%%%%%%%%%%%%%%%%%%%%

%%%%%%%%%%%%%%%%%%%%%%%%%%%%%%%%%%%%%%%%%%%
101Q  $\leq m \leq$ 120Q & &
368 & 34 & 334  & &
81 & 0 & 81 & &
12,925 & 59 & 12,866  & &
13,374 & 93 & 13,281 \\ \addlinespace[5pt]
%%%%%%%%%%%%%%%%%%%%%%%%%%%%%%%%%%%%%%%%%%%

%%%%%%%%%%%%%%%%%%%%%%%%%%%%%%%%%%%%%%%%%%%
121Q  $\leq m \leq$ 200Q & &
171 & 49 & 122 & &
18 & 0 & 18 & &
2,767 & 215 & 2,552 & &
2,956 & 264 & 2,692 \\ \addlinespace[5pt]
%%%%%%%%%%%%%%%%%%%%%%%%%%%%%%%%%%%%%%%%%%%

\addlinespace[5pt]

%%%%%%%%%%%%%%%%%%%%%%%%%%%%%%%%%%%%%%%%%%%
Total & &
78,632 & 2,804 & 75,828 & & 
85,714 & 1,984 & 83,730 & &
260,973 & 5,401 & 258,572  & &
430,504 & 10,289 & 420,315 \\ \addlinespace[5pt]
%%%%%%%%%%%%%%%%%%%%%%%%%%%%%%%%%%%%%%%%%%%

\addlinespace[10pt]

& & \multicolumn{15}{c}{Zombie-Definition: $Z^{BROAD}$} \\ 
\cmidrule(lr){3-17}
& & \multicolumn{3}{c}{Bank/Term Loans} & & \multicolumn{3}{c}{Revolving Credit Facility} & & \multicolumn{3}{c}{Bonds and Notes} & & \multicolumn{3}{c}{Total}  \\ 
\cmidrule(lr){3-5} \cmidrule(lr){7-9} \cmidrule(lr){11-13} \cmidrule(lr){15-17} 
& &All & Zombies & Non-Zombies & & All & Zombies & Non-Zombies & & All & Zombies & Non-Zombies & & All & Zombies & Non-Zombies \\ 
\cmidrule(lr){3-3} \cmidrule(lr){4-4} \cmidrule(lr){5-5} \cmidrule(lr){7-7} \cmidrule(lr){8-8}  \cmidrule(lr){9-9} \cmidrule(lr){11-11} \cmidrule(lr){12-12} \cmidrule(lr){13-13} \cmidrule(lr){15-15} \cmidrule(lr){16-16} \cmidrule(lr){17-17} 

%%%%%%%%%%%%%%%%%%%%%%%%%%%%%%%%%%%%%%%%%%%
1Q  $\leq m \leq$ 4Q  & &
14,174 & 1,583 & 12,591 & &
17,834 & 1,349 & 16,485 & &
34,302 & 6,449 & 27,853  & &
67,198 & 9,381 & 57,817 \\ \addlinespace[5pt]
%%%%%%%%%%%%%%%%%%%%%%%%%%%%%%%%%%%%%%%%%%%

%%%%%%%%%%%%%%%%%%%%%%%%%%%%%%%%%%%%%%%%%%%
5Q  $\leq m \leq$ 8Q & &
10,831 & 1,365 & 9,466 & &
12,876 & 1,022 & 11,854 & &
28,212 & 3,834 & 24,378 & &
52,230 & 6,221 & 46,009 \\ \addlinespace[5pt]
%%%%%%%%%%%%%%%%%%%%%%%%%%%%%%%%%%%%%%%%%%%

%%%%%%%%%%%%%%%%%%%%%%%%%%%%%%%%%%%%%%%%%%%
9Q  $\leq m \leq$ 20Q & &
30,425 & 2,457 & 27,968 & &
33,630 & 1,443 & 32,187 & &
75,140 & 5,979 & 69,161 & &
139,830 & 9,879 & 129,951 \\ \addlinespace[5pt]
%%%%%%%%%%%%%%%%%%%%%%%%%%%%%%%%%%%%%%%%%%%

%%%%%%%%%%%%%%%%%%%%%%%%%%%%%%%%%%%%%%%%%%%
21Q  $\leq m \leq$ 40Q & &
18,284 & 999 & 17,285 & &
17,673 & 461 & 17,212  & &
75,449 & 2,191 & 73,258  & &
111,717 & 3,651 & 108,066 \\ \addlinespace[5pt]
%%%%%%%%%%%%%%%%%%%%%%%%%%%%%%%%%%%%%%%%%%%

%%%%%%%%%%%%%%%%%%%%%%%%%%%%%%%%%%%%%%%%%%%
41Q  $\leq m \leq$ 100Q & &
4,379 & 264 & 4,115 & &
3,602 & 77 & 3,525 & &
35,178 & 711 & 34,467 & &
43,199 & 1,052 & 42,147 \\ \addlinespace[5pt]
%%%%%%%%%%%%%%%%%%%%%%%%%%%%%%%%%%%%%%%%%%%

%%%%%%%%%%%%%%%%%%%%%%%%%%%%%%%%%%%%%%%%%%%
101Q  $\leq m \leq$ 120Q & &
368 & 56 & 312 & &
81 & 0 & 81 & &
12,925 & 87 & 12,838  & &
13,374 & 143 & 13,231 \\ \addlinespace[5pt]
%%%%%%%%%%%%%%%%%%%%%%%%%%%%%%%%%%%%%%%%%%%

%%%%%%%%%%%%%%%%%%%%%%%%%%%%%%%%%%%%%%%%%%%
121Q  $\leq m \leq$ 200Q & &
171 & 52 & 119 & &
18 & 0 & 18 & &
2,767 & 220 & 2,547 & &
2,956 & 272 & 2,684 \\ \addlinespace[5pt]
%%%%%%%%%%%%%%%%%%%%%%%%%%%%%%%%%%%%%%%%%%%

\addlinespace[5pt]

%%%%%%%%%%%%%%%%%%%%%%%%%%%%%%%%%%%%%%%%%%%
Total & &
78,632 & 6,776 & 71,856 & & 
85,714 & 4,352 & 81,362 & &
263,973 & 19,471 & 244,502  & &
430,504 & 30,599 & 399,905 \\ \addlinespace[5pt]
%%%%%%%%%%%%%%%%%%%%%%%%%%%%%%%%%%%%%%%%%%%

\hline

\end{tabular}
%\end{center}
\begin{tablenotes}
\scriptsize 
\item[] Notes: We show the total number of newly reported debt obligations in company filings in Compustat's Capital-IQ database in the years 2002-2020. The left-most columns shows the different maturity bins.
\end{tablenotes}
\end{threeparttable}
\end{table}
\end{landscape}

%\vspace{2cm}
\section{``Zombies'' -- Non-Viable Firms} \label{sec:ZombieDef}
Even though the issue of zombie prevalence has been discussed in previous studies already, the literature does not provide a uniform definition of a ``zombie-firm'', but deploys various classification schemes -- also depending on data availability.\footnote{For definitions different from the ones used in this paper, see for example \cite{CaballeroHoshiKashyap2008}, \cite{AcharyaEtAl2019}, \cite{AndrewsPetroulakis2019}.} In this section, we will discuss two prominent definitions, which require a small number of financial variables and which are compatible with our data set.

\subsection{The Broader Definition} 
Our first approach to separate non-viable firms ($Z_{i,t}$) from their healthy peers ($NZ_{i,t}$), follows \cite{McGowanAndrewsMillot2018,banerjee2018rise,AndrewsPetroulakis2019} and define a zombie as a firm that (I) has reported an Interest Coverage Ratio ($ICR_{i,t} = XINT_{i,t} / EBITDA_{i,t}$)\footnote{\textit{XINT} = Interest and Related Expense Total and  \textit{EBITDA} = Earnings Before Interest, Taxes, Depreciation, and Amortization} larger than one for three consecutive years, and (II) is at least 10 years of age.
We refer to firms, identified as zombies according to this definition, as $Z^{BROAD}$.
\cite{banerjee2018rise} raise the concern that this definition is not stringent enough to single out those companies, which experience temporary difficulties, but which are generally perceived to have encouraging growth prospects. In their sample, the median Tobin's Q -- a proxy for expected future profitability as perceived by stock market participants -- of zombies is higher than the non-zombie counterpart.

\subsection{A Narrower Definition} 
\cite{banerjee2018rise} therefore extend the previous (broader) zombie-definition ($Z^{BROAD}_t$) by accounting for a firm's growth potential. 
Recall that our prevailing definition of firm $i$ being a zombie is based on two requirements: (I) a firm's ICR over the past three years has not fallen below 1 and (II) the firm is at least 10 years of age. \cite{banerjee2018rise} augment this definition by requiring that (III) a firm's \textit{Tobin's Q} in a given year $t$ ranges below the industry-median.\footnote{We define an industry via two-digit NAICS codes.} Let us define the set of companies, meeting all three requirements in a given year $t$, as $Z^{NAR.X}$.

\noindent Nonetheless, this more conservative definition does not come without costs, and if adopted without further considerations, introduces non-negligible distortions.
The first issue arising from $Z^{NAR.X}$ is a potential decline in the total number of companies. This stems from the sparsity of reported \textit{Tobin's Q}, which shrinks the set of companies being eligible for classification. This reduced set of observations also alters the characteristics of zombies and non-zombies. 
Table \ref{tab:SumStat_Compustat_ZvsZ2} compares company fundamentals for the two zombie-definitions $Z^{BROAD}$ and $Z^{NAR.X}$. Under the narrow definition, the median zombie is more productive, larger, less indebted, and more profitable than zombies under the broader measure.

\begin{landscape}
    \begin{table}[h]
    %\begin{center}
    \centering
    \begin{threeparttable}
    \captionsetup{justification=centering}
    \scriptsize
    \vspace{-1cm}
\caption{\label{tab:SumStat_Compustat_ZvsZ2} Summary Statistics -- Full Sample: 2002-2020}

\begin{tabular}{l cccc c cccc c c} \hline

& \multicolumn{4}{c}{Mean} & & \multicolumn{4}{c}{Median} & & \multirow{2}{*}{Units} \\ 
\cmidrule(lr){2-5} \cmidrule(lr){7-10} 

& All$^\text{BROAD}$ & All$^\text{NAR.X}$ & $Z^{BROAD}$ & $Z^{NAR.X}$ & & All$^\text{BROAD}$ & All$^\text{NAR.X}$ & $Z^{BROAD}$ & $Z^{NAR.X}$ & &  \\
\cmidrule(lr){2-2}  \cmidrule(lr){3-3}  \cmidrule(lr){4-4}  \cmidrule(lr){5-5}  \cmidrule(lr){7-7} \cmidrule(lr){8-8}  \cmidrule(lr){9-9} \cmidrule(lr){10-10} \cmidrule(lr){12-12}

% %%%%%%%%%%%%%%%%%%%%%%%%%%%%%%%%%%%%%%%%%%%
% Labor Productivity & 
% 0.01 & 0.01 & -0.21 & -0.14 & &
% 0.04 & 0.04 & -0.07 & -0.01 & &
% Mill. USD / Employee \\ \addlinespace[5pt]
% %%%%%%%%%%%%%%%%%%%%%%%%%%%%%%%%%%%%%%%%%%%

%%%%%%%%%%%%%%%%%%%%%%%%%%%%%%%%%%%%%%%%%%%
TFP & 
0.03 & 0.04 & -0.29 & -0.19 & &
0.10 & 0.10 & 0.08 & 0.10 & &
 \\ \addlinespace[5pt]
%%%%%%%%%%%%%%%%%%%%%%%%%%%%%%%%%%%%%%%%%%%

%%%%%%%%%%%%%%%%%%%%%%%%%%%%%%%%%%%%%%%%%%%
Assets & 
2.50 & 2.57 & 0.21 & 0.29 & &
0.17 & 0.17 & 0.01 & 0.03 & &
Bill. USD \\ \addlinespace[5pt]
%%%%%%%%%%%%%%%%%%%%%%%%%%%%%%%%%%%%%%%%%%%

%%%%%%%%%%%%%%%%%%%%%%%%%%%%%%%%%%%%%%%%%%%
Sales & 
2.04 & 2.12 & 0.19 & 0.16 & &
0.14 & 0.14 & 0.00 & 0.01 & &
Bill. USD \\ \addlinespace[5pt]
%%%%%%%%%%%%%%%%%%%%%%%%%%%%%%%%%%%%%%%%%%%

%%%%%%%%%%%%%%%%%%%%%%%%%%%%%%%%%%%%%%%%%%%
(Book) Leverage & 
2.36 & 2.52 & 11.44 & 0.23 & &
0.22 & 0.21 & 0.23 & 0.13 & &
 \\ \addlinespace[5pt]
%%%%%%%%%%%%%%%%%%%%%%%%%%%%%%%%%%%%%%%%%%%

%%%%%%%%%%%%%%%%%%%%%%%%%%%%%%%%%%%%%%%%%%%
Asset Tangibility & 
0.24 & 0.24 & 0.18 & 0.19 & &
0.15 & 0.14 & 0.07 & 0.08 & &
 \\ \addlinespace[5pt]
%%%%%%%%%%%%%%%%%%%%%%%%%%%%%%%%%%%%%%%%%%%

%%%%%%%%%%%%%%%%%%%%%%%%%%%%%%%%%%%%%%%%%%%
CapX / Assets & 
0.06 & 0.06 & 0.05 & 0.03 & &
0.02 & 0.02 & 0.01 & 0.01 & &
 \\ \addlinespace[5pt]
%%%%%%%%%%%%%%%%%%%%%%%%%%%%%%%%%%%%%%%%%%%

%%%%%%%%%%%%%%%%%%%%%%%%%%%%%%%%%%%%%%%%%%%
ROA & 
-6.34 & -6.97 & -23.36 & -0.29 & &
0.00 & 0.01 & -0.40 & -0.18 & &
 \\ \addlinespace[5pt]
%%%%%%%%%%%%%%%%%%%%%%%%%%%%%%%%%%%%%%%%%%%

%%%%%%%%%%%%%%%%%%%%%%%%%%%%%%%%%%%%%%%%%%%
Value Added & 
643.29 & 653.76 & 90.63 & 27.43 & &
40.34 & 37.39 & -1.32 & -0.31 & &
Mill. USD \\ \addlinespace[5pt]
%%%%%%%%%%%%%%%%%%%%%%%%%%%%%%%%%%%%%%%%%%%

%%%%%%%%%%%%%%%%%%%%%%%%%%%%%%%%%%%%%%%%%%%
Age & 
16.97 & 18.30 & 19.39 & 20.84 & &
12.00 & 14.00 & 17.00 & 18.00 & &
Years \\ \addlinespace[5pt]
%%%%%%%%%%%%%%%%%%%%%%%%%%%%%%%%%%%%%%%%%%%

%%%%%%%%%%%%%%%%%%%%%%%%%%%%%%%%%%%%%%%%%%%
Employees ($\times$ 10$^3$) & 
9.01 & 8.94 & 1.57 & 0.92 & &
0.70 & 0.66 & 0.05 & 0.09 & &
 \\ \addlinespace[5pt]
%%%%%%%%%%%%%%%%%%%%%%%%%%%%%%%%%%%%%%%%%%%

%%%%%%%%%%%%%%%%%%%%%%%%%%%%%%%%%%%%%%%%%%%
Tobin's-Q & 
84.69 & 84.69 & 59.54 & 1.20 & &
1.70 & 1.70 & 2.78 & 1.13 & &
 \\ \addlinespace[5pt]
%%%%%%%%%%%%%%%%%%%%%%%%%%%%%%%%%%%%%%%%%%%

\addlinespace[3pt]
\cmidrule(lr){1-10}
\addlinespace[3pt]

& \multicolumn{4}{c}{IQR} & & \multicolumn{4}{c}{SD} & & \multirow{2}{*}{Units} \\ 
\cmidrule(lr){2-5} \cmidrule(lr){7-10} 

& All$^\text{BROAD}$ & All$^\text{NAR.X}$ & $Z^{BROAD}$ & $Z^{NAR.X}$ & & All$^\text{BROAD}$ & All$^\text{NAR.X}$ & $Z^{BROAD}$ & $Z^{NAR.X}$ & &  \\
\cmidrule(lr){2-2}  \cmidrule(lr){3-3}  \cmidrule(lr){4-4}  \cmidrule(lr){5-5}  \cmidrule(lr){7-7} \cmidrule(lr){8-8}  \cmidrule(lr){9-9} \cmidrule(lr){10-10} \cmidrule(lr){12-12}

% %%%%%%%%%%%%%%%%%%%%%%%%%%%%%%%%%%%%%%%%%%%
% Labor Productivity & 
% 0.07 & 0.07 & 0.25 & 0.19 & &
% 1.04 & 1.01 & 0.51 & 0.42 & &
% Mill. USD / Employee \\ \addlinespace[5pt]
% %%%%%%%%%%%%%%%%%%%%%%%%%%%%%%%%%%%%%%%%%%%

%%%%%%%%%%%%%%%%%%%%%%%%%%%%%%%%%%%%%%%%%%%
TFP & 
0.52 & 0.52 & 0.97 & 0.70 & &
0.93 & 0.93 & 1.51 & 1.25 & &
 \\ \addlinespace[5pt]
%%%%%%%%%%%%%%%%%%%%%%%%%%%%%%%%%%%%%%%%%%%

%%%%%%%%%%%%%%%%%%%%%%%%%%%%%%%%%%%%%%%%%%%
Assets & 
1.00 & 1.03 & 0.05 & 0.10 & &
12.67 & 13.18 & 1.40 & 1.38 & &
Bill. USD \\ \addlinespace[5pt]
%%%%%%%%%%%%%%%%%%%%%%%%%%%%%%%%%%%%%%%%%%%

%%%%%%%%%%%%%%%%%%%%%%%%%%%%%%%%%%%%%%%%%%%
Sales & 
0.86 & 0.88 & 0.03 & 0.06 & &
10.47 & 10.92 & 2.52 & 0.88 & &
Bill. USD \\ \addlinespace[5pt]
%%%%%%%%%%%%%%%%%%%%%%%%%%%%%%%%%%%%%%%%%%%

%%%%%%%%%%%%%%%%%%%%%%%%%%%%%%%%%%%%%%%%%%%
(Book) Leverage & 
0.42 & 0.40 & 0.74 & 0.37 & &
75.44 & 79.51 & 206.81 & 0.28 & &
 \\ \addlinespace[5pt]
%%%%%%%%%%%%%%%%%%%%%%%%%%%%%%%%%%%%%%%%%%%

%%%%%%%%%%%%%%%%%%%%%%%%%%%%%%%%%%%%%%%%%%%
Asset Tangibility & 
0.31 & 0.31 & 0.20 & 0.24 & &
0.25 & 0.25 & 0.24 & 0.25 & &
 \\ \addlinespace[5pt]
%%%%%%%%%%%%%%%%%%%%%%%%%%%%%%%%%%%%%%%%%%%

%%%%%%%%%%%%%%%%%%%%%%%%%%%%%%%%%%%%%%%%%%%
CapX / Assets & 
0.04 & 0.04 & 0.03 & 0.02 & &
0.44 & 0.46 & 0.55 & 0.10 & &
 \\ \addlinespace[5pt]
%%%%%%%%%%%%%%%%%%%%%%%%%%%%%%%%%%%%%%%%%%%

%%%%%%%%%%%%%%%%%%%%%%%%%%%%%%%%%%%%%%%%%%%
ROA & 
0.29 & 0.29 & 0.94 & 0.32 & &
499.50 & 527.08 & 1472.09 & 0.59 & &
 \\ \addlinespace[5pt]
%%%%%%%%%%%%%%%%%%%%%%%%%%%%%%%%%%%%%%%%%%%

%%%%%%%%%%%%%%%%%%%%%%%%%%%%%%%%%%%%%%%%%%%
Value Added & 
283.55 & 278.52 & 7.20 & 9.08 & &
3085.94 & 3042.25 & 1995.76 & 320.73 & &
Mill. USD \\ \addlinespace[5pt]
%%%%%%%%%%%%%%%%%%%%%%%%%%%%%%%%%%%%%%%%%%%

%%%%%%%%%%%%%%%%%%%%%%%%%%%%%%%%%%%%%%%%%%%
Age & 
19.00 & 18.00 & 10.00 & 12.00 & &
15.46 & 15.45 & 8.99 & 10.71 & &
Years \\ \addlinespace[5pt]
%%%%%%%%%%%%%%%%%%%%%%%%%%%%%%%%%%%%%%%%%%%

%%%%%%%%%%%%%%%%%%%%%%%%%%%%%%%%%%%%%%%%%%%
Employees ($\times$ 10$^3$) & 
4.38 & 4.30 & 0.16 & 0.28 & &
46.75 & 46.50 & 26.54 & 6.49 & &
 \\ \addlinespace[5pt]
%%%%%%%%%%%%%%%%%%%%%%%%%%%%%%%%%%%%%%%%%%%

%%%%%%%%%%%%%%%%%%%%%%%%%%%%%%%%%%%%%%%%%%%
Tobin's-Q & 
1.94 & 1.94 & 5.39 & 0.61 & &
3808.22 & 3808.22 & 827.25 & 253.24 & &
 \\ \addlinespace[5pt]
%%%%%%%%%%%%%%%%%%%%%%%%%%%%%%%%%%%%%%%%%%%

\hline

\end{tabular}
%\end{center}
\begin{tablenotes}
\scriptsize 
\item[]  Notes: For each variable we calculate the mean, median, IQR, and SD, along both the time- and cross-sectional dimension.
See the Appendix for the computation of each variable.
\end{tablenotes}
\end{threeparttable}
\end{table}
\end{landscape}

\noindent An explanation for this puzzle can be found in \cite{BStoffiRiccaboniRungi2020}, who detect a positive correlation between a firm's failure and it holding back financial information. 
Similar to \cite{BStoffiRiccaboniRungi2020}, we compute a binary variable for a firm's exit ($E_{i,t}$), which equals 1 in the year prior to its liquidation, and zero otherwise. Another binary variable ($V^B_{i,t}$), is set to 1 in year $t$ if $E_{i,t} = 1$ and if a certain financial variable of interest ($V$) --  e.g. \textit{Tobin's Q} or \textit{Interest Expenses} -- was missing at least once in the current or preceding two years. Using $V = \{\text{Tobin's Q}, \; \text{Interest Expenses}\}$ we get a correlation of $Cor\left(E,V^B\right) = 0.55$ in the case of \textit{Tobin's Q} and $Cor\left(E,V^B\right) = 0.48$ in the case of \textit{Interest Expenses}. These numbers suggest that firms are either more reluctant or less capable of reporting information necessary to calculate their \textit{Tobin's Q}
%\footnote{Variables to compute \textit{Tobin's Q} also enter the formula for the \textit{market-to-book} ratio, which renders this variable as a control in our regressions inappropriate.}, 
than reporting \textit{Interest Expenses} in the three years prior to their passing. Thus, conditioning our definition of zombies on the more restrictive definition, which includes \textit{Tobin's Q}, mechanically excludes not only potential zombies but literally \textit{non-viable} firms from the sample.%\footnote{Therefore we also keep our firm controls to a minimum while trying to closely follow the established literature.}

\noindent Despite its reasonable theoretical underpinnings, the inclusion of \textit{Tobin's Q} as an additional criterion for classifying zombies, ultimately falls short of capturing the most distressed firms. 
%To overcome this selection bias, one could follow along two strategies: a first option is to apply a broad definition, which requires only little firm-level reporting, but can be further structured by differentiating among subsets of zombies, such as \textit{recovering} and \textit{remaining} zombies. The second, and more comprehensive, approach is to account for patterns in missing financial information \citep{BStoffiRiccaboniRungi2020}. %using machine learning algorithms such as Bayesian Additive Regression Tree with Missing Incorporated in Attributes (BART-MIA).
%To overcome the above named issue of the narrow definition falling short of capturing firms, which do not report \textit{Tobin's Q}, 
We therefore modify $Z^{NAR.X}$ by classifying a firm as a zombie in year $t$, if it either complies with all three requirements of the narrow definition or if it fails to report \textit{Tobin's Q} in year $t$.
We call the set of firms complying with this extended narrower definition $Z^{NAR}$.
Table \ref{tab:SumStat_Compustat_v3} shows summary statistics for $Z^{NAR}$ for several company fundamentals over the period 2002 and 2020. Zombies are on average less productive than their viable counterparts, smaller -- both in terms of assets and employees -- and operate with less leverage.

%\begin{landscape}
    \begin{table}[h!]
	\centering
	\begin{threeparttable}
    %\begin{center}
    \captionsetup{justification=centering}
    \scriptsize
    \vspace{1cm}
\caption{\label{tab:SumStat_Compustat_v3} Summary Statistics -- Full Sample: 2002-2020}

\begin{tabular}{l ccc c ccc c c} \hline

& \multicolumn{3}{c}{Mean} & & \multicolumn{3}{c}{Median} & & \multirow{2}{*}{Units} \\ 
\cmidrule(lr){2-4} \cmidrule(lr){6-8} 

& All & $Z^{NAR}$ & Non-Zombies & & All & $Z^{NAR}$ & Non-Zombies  \\
\cmidrule(lr){2-2}  \cmidrule(lr){3-3}  \cmidrule(lr){4-4}  \cmidrule(lr){6-6}  \cmidrule(lr){7-7} \cmidrule(lr){8-8}  \cmidrule(lr){10-10}

% %%%%%%%%%%%%%%%%%%%%%%%%%%%%%%%%%%%%%%%%%%%
% Labor Productivity & 
% 0.02 & -0.14 & 0.02 & &
% 0.04 & -0.01 & 0.05   & &
% Mill. USD / Employee \\ \addlinespace[5pt]
% %%%%%%%%%%%%%%%%%%%%%%%%%%%%%%%%%%%%%%%%%%%

%%%%%%%%%%%%%%%%%%%%%%%%%%%%%%%%%%%%%%%%%%%
TFP & 
0.03 & -0.18 & 0.04 & &
 0.10 & 0.10 & 0.10   & &
 \\ \addlinespace[5pt]
%%%%%%%%%%%%%%%%%%%%%%%%%%%%%%%%%%%%%%%%%%%

%%%%%%%%%%%%%%%%%%%%%%%%%%%%%%%%%%%%%%%%%%%
Assets & 
2.50 & 0.40 & 2.57 & &
0.17 & 0.03 & 0.19   & &
Bill. USD \\ \addlinespace[5pt]
%%%%%%%%%%%%%%%%%%%%%%%%%%%%%%%%%%%%%%%%%%%

%%%%%%%%%%%%%%%%%%%%%%%%%%%%%%%%%%%%%%%%%%%
Sales & 
2.04 & 0.43 & 2.10 & &
0.14 & 0.01 & 0.15  & &
Bill. USD \\ \addlinespace[5pt]
%%%%%%%%%%%%%%%%%%%%%%%%%%%%%%%%%%%%%%%%%%%

%%%%%%%%%%%%%%%%%%%%%%%%%%%%%%%%%%%%%%%%%%%
(Book) Leverage & 
2.36 & 0.70 & 2.42 & &
0.22 & 0.13 & 0.22   & &
 \\ \addlinespace[5pt]
%%%%%%%%%%%%%%%%%%%%%%%%%%%%%%%%%%%%%%%%%%%

%%%%%%%%%%%%%%%%%%%%%%%%%%%%%%%%%%%%%%%%%%%
Asset Tangibility & 
0.24 & 0.20 & 0.24 & &
0.15 & 0.08 & 0.15  & &
 \\ \addlinespace[5pt]
%%%%%%%%%%%%%%%%%%%%%%%%%%%%%%%%%%%%%%%%%%%

%%%%%%%%%%%%%%%%%%%%%%%%%%%%%%%%%%%%%%%%%%%
CapX / Assets & 
0.06 & 0.03 & 0.06 & &
0.02 & 0.01 & 0.03  & &
 \\ \addlinespace[5pt]
%%%%%%%%%%%%%%%%%%%%%%%%%%%%%%%%%%%%%%%%%%%

%%%%%%%%%%%%%%%%%%%%%%%%%%%%%%%%%%%%%%%%%%%
ROA & 
-6.34 & -0.02 & -6.57 & &
0.00 & -0.18 & 0.01  & &
 \\ \addlinespace[5pt]
%%%%%%%%%%%%%%%%%%%%%%%%%%%%%%%%%%%%%%%%%%%

%%%%%%%%%%%%%%%%%%%%%%%%%%%%%%%%%%%%%%%%%%%
Value Added & 
643.29 & 250.38 & 658.13 & &
40.34 & -0.26 & 46.67   & & 
Mill. USD \\ \addlinespace[5pt]
%%%%%%%%%%%%%%%%%%%%%%%%%%%%%%%%%%%%%%%%%%%

%%%%%%%%%%%%%%%%%%%%%%%%%%%%%%%%%%%%%%%%%%%
Age & 
16.97 & 20.87 & 16.83 & &
12.00 & 18.00 & 12.00   & &
Years \\ \addlinespace[5pt]
%%%%%%%%%%%%%%%%%%%%%%%%%%%%%%%%%%%%%%%%%%%

%%%%%%%%%%%%%%%%%%%%%%%%%%%%%%%%%%%%%%%%%%%
Employees ($\times$ 10$^3$) & 
9.01 & 3.86 & 9.20 & &
0.70 & 0.09 & 0.78   & &
 \\ \addlinespace[5pt]
%%%%%%%%%%%%%%%%%%%%%%%%%%%%%%%%%%%%%%%%%%%

%%%%%%%%%%%%%%%%%%%%%%%%%%%%%%%%%%%%%%%%%%%
Tobin's-Q & 
84.69 & 1.20 & 87.97 & &
1.70 & 1.13 & 1.74   & &
 \\ \addlinespace[5pt]
%%%%%%%%%%%%%%%%%%%%%%%%%%%%%%%%%%%%%%%%%%%

\addlinespace[3pt]
\cmidrule(lr){1-10}
\addlinespace[3pt]

& \multicolumn{3}{c}{IQR} & & \multicolumn{3}{c}{SD}  & & \multirow{2}{*}{Units} \\ 
\cmidrule(lr){2-4} \cmidrule(lr){6-8} 

& All & $Z^{NAR}$ & Non-Zombies & & All & $Z^{NAR}$ & Non-Zombies  \\
\cmidrule(lr){2-2}  \cmidrule(lr){3-3}  \cmidrule(lr){4-4}  \cmidrule(lr){6-6}  \cmidrule(lr){7-7} \cmidrule(lr){8-8} \cmidrule(lr){10-10}

% %%%%%%%%%%%%%%%%%%%%%%%%%%%%%%%%%%%%%%%%%%%
% Labor Productivity & 
% 0.07 & 0.19 & 0.07 & &
% 1.04 & 0.43 & 1.05  & &
% Mill. USD / Employee \\ \addlinespace[5pt]
% %%%%%%%%%%%%%%%%%%%%%%%%%%%%%%%%%%%%%%%%%%%

%%%%%%%%%%%%%%%%%%%%%%%%%%%%%%%%%%%%%%%%%%%
TFP & 
0.52 & 0.68 & 0.52 & &
0.93 & 1.24 & 0.92  & &
 \\ \addlinespace[5pt]
%%%%%%%%%%%%%%%%%%%%%%%%%%%%%%%%%%%%%%%%%%%

%%%%%%%%%%%%%%%%%%%%%%%%%%%%%%%%%%%%%%%%%%%
Assets & 
1.00 & 0.10 & 1.06  & &
12.67 & 2.12 & 12.89  & &
Bill. USD \\ \addlinespace[5pt]
%%%%%%%%%%%%%%%%%%%%%%%%%%%%%%%%%%%%%%%%%%%

%%%%%%%%%%%%%%%%%%%%%%%%%%%%%%%%%%%%%%%%%%%
Sales & 
0.86 & 0.06 & 0.91 & &
10.47 & 4.11 & 10.62  & &
Bill. USD \\ \addlinespace[5pt]
%%%%%%%%%%%%%%%%%%%%%%%%%%%%%%%%%%%%%%%%%%%

%%%%%%%%%%%%%%%%%%%%%%%%%%%%%%%%%%%%%%%%%%%
(Book) Leverage & 
0.42 & 0.40 & 0.42 & &
75.44 & 15.84 & 76.73   & &
 \\ \addlinespace[5pt]
%%%%%%%%%%%%%%%%%%%%%%%%%%%%%%%%%%%%%%%%%%%

%%%%%%%%%%%%%%%%%%%%%%%%%%%%%%%%%%%%%%%%%%%
Asset Tangibility & 
0.31 & 0.24 & 0.31 & &
0.25 & 0.26 & 0.25   & &
 \\ \addlinespace[5pt]
%%%%%%%%%%%%%%%%%%%%%%%%%%%%%%%%%%%%%%%%%%%

%%%%%%%%%%%%%%%%%%%%%%%%%%%%%%%%%%%%%%%%%%%
CapX / Assets & 
0.04 & 0.02 & 0.05 & &
0.44 & 0.10 & 0.44   & &
 \\ \addlinespace[5pt]
%%%%%%%%%%%%%%%%%%%%%%%%%%%%%%%%%%%%%%%%%%%

%%%%%%%%%%%%%%%%%%%%%%%%%%%%%%%%%%%%%%%%%%%
ROA & 
0.29 & 0.32 & 0.28 & &
499.50 & 17.81 & 508.45 & &
 \\ \addlinespace[5pt]
%%%%%%%%%%%%%%%%%%%%%%%%%%%%%%%%%%%%%%%%%%%

%%%%%%%%%%%%%%%%%%%%%%%%%%%%%%%%%%%%%%%%%%%
Value Added & 
283.55 & 9.27 & 301.74 & &
3085.94 & 3269.36 & 3077.34 & & 
Mill. USD \\ \addlinespace[5pt]
%%%%%%%%%%%%%%%%%%%%%%%%%%%%%%%%%%%%%%%%%%%

%%%%%%%%%%%%%%%%%%%%%%%%%%%%%%%%%%%%%%%%%%%
Age & 
19.00 & 12.00 & 19.00 & &
15.46 & 10.64 & 15.59  & &
Years \\ \addlinespace[5pt]
%%%%%%%%%%%%%%%%%%%%%%%%%%%%%%%%%%%%%%%%%%%

%%%%%%%%%%%%%%%%%%%%%%%%%%%%%%%%%%%%%%%%%%%
Employees ($\times$ 10$^3$) & 
4.38 & 0.30 & 4.65 & &
46.75 & 43.47 & 46.85  & &
 \\ \addlinespace[5pt]
%%%%%%%%%%%%%%%%%%%%%%%%%%%%%%%%%%%%%%%%%%%

%%%%%%%%%%%%%%%%%%%%%%%%%%%%%%%%%%%%%%%%%%%
Tobin's-Q & 
1.94 & 0.61 & 2.03 & &
3808.22 & 0.48 & 3882.26  & &
 \\ \addlinespace[5pt]
%%%%%%%%%%%%%%%%%%%%%%%%%%%%%%%%%%%%%%%%%%%

\hline

\end{tabular}
%\end{center}
\begin{tablenotes}
\scriptsize 
\item[]  Notes: For each variable we calculate the mean, median, IQR, and SD, along both the time- and cross-sectional dimension.
See the Appendix for the computation of each variable.
\end{tablenotes}
\end{threeparttable}
\end{table}
%\end{landscape}

\begin{figure}[h!]
	\centering
    %\begin{center}
    \caption{\footnotesize{Zombie Prevalence under Different Zombie-Definitions}} \label{fig:ZvsZ3}
    \begin{threeparttable}
        \includegraphics[width=0.75\textwidth, trim = 0mm 60mm 0mm 65mm, clip]{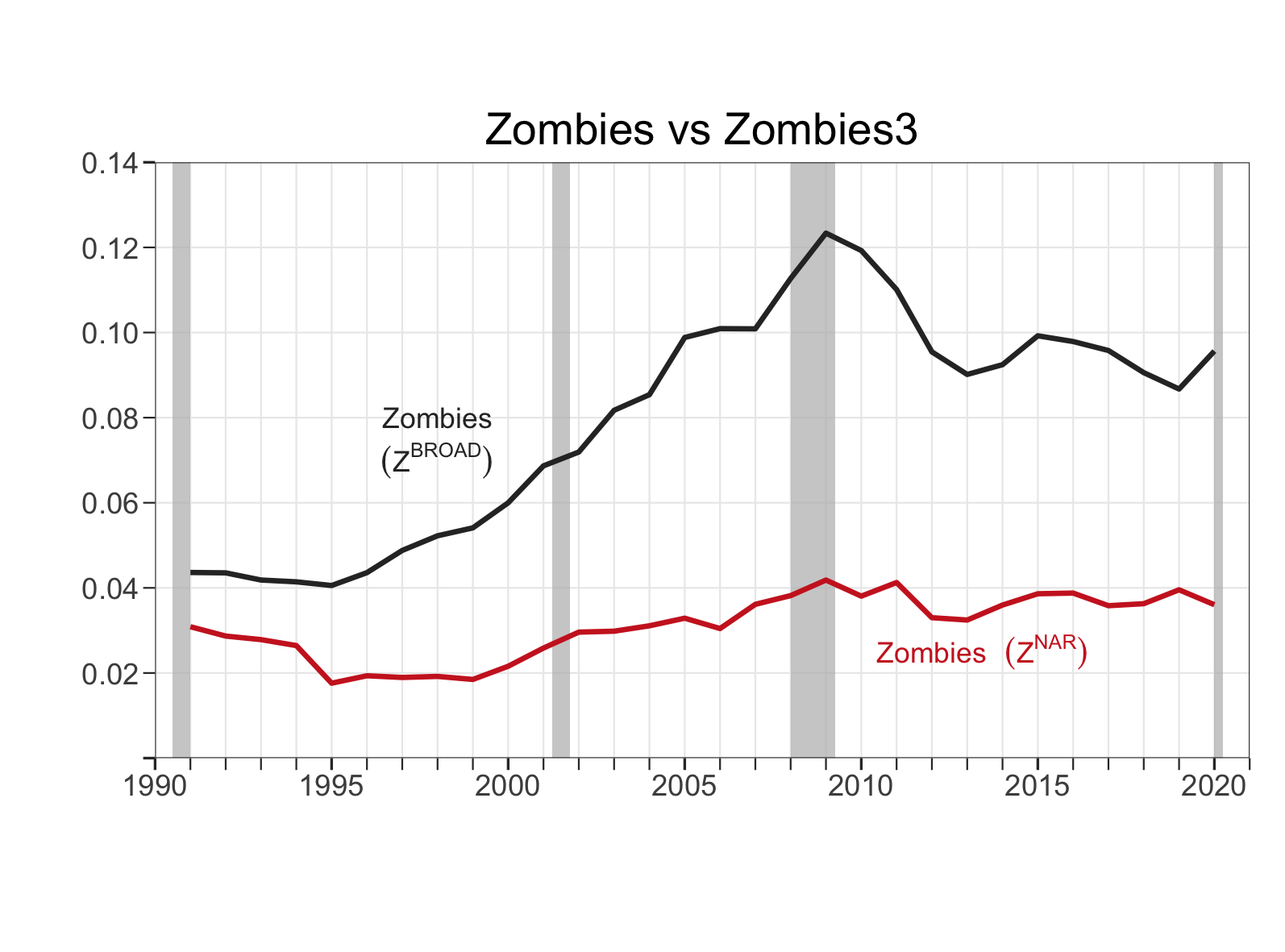}
    %\end{center}
    \vskip 5pt
    \begin{tablenotes}
    \scriptsize
\item[]  \hspace{3cm} Notes: Shaded areas mark NBER recessions.
 \end{tablenotes}
 \end{threeparttable}
\end{figure}

Figure \ref{fig:ZvsZ3} provides an overview of zombie-prevalence since the early 1990s under the two zombie definitions $Z^{BROAD}$ and $Z^{NAR}$. The differences are remarkable, peaking at about eight percentage points at the end of the Great Recession. Though, both definitions share a steady upward trend starting in the mid 1990s and ending during the financial crisis of 2007 through 2009, and a decline thereafter. 

\clearpage
\subsection{``Zombie''-Lending} \label{sec:zombies_zombielending}
%So far, many empirical studies \citep{schivardi2017credit,McGowanAndrewsMillot2018,AndrewsPetroulakis2019,schivardi2020identifying} have evaluated the effect of zombie-prevalence and/or zombie-lending in the Euro Area. This study shifts the focus onto the United States. 

\noindent Our empirical analysis focuses mainly on the implications of zombie-lending for the performance of non-zombies. Zombie-lending can occur in the form of increased credit supply and/or reduced costs of funding \citep{AcharyaLenzuWang2021}. While the latter has attracted heightened attention in the literature\footnote{See \cite{AcharyaLenzuWang2021} for a more comprehensive overview of related studies.}, our analysis targets the former, i.e. supply of credit to non-viable firms. 

\noindent As outlined in Section \ref{sec:data_debt}, CapIQ allows us to differentiate among different types of debt and different maturities. Sticking with our two zombie-definitions, Table \ref{tab:SummStats_CapIQ_maturities} provides an overview of the overall number of new $BL$, $RC$, and $BN$ filings, split into several maturity bins. It is immediately apparent that the number of observations associated with zombie-lending decreases significantly when shifting from the broader zombie-definition $Z^{BROAD}$ in the lower panel to $Z^{NAR}$, shown in the upper panel. Overall, $BN$ also make up the bulk of zombie-lending with a little over 52\% under $Z^{NAR}$ and almost 64\% under $Z^{BROAD}$ of total observations reported by zombies. In terms of bank-credit, zombies -- other than their non-zombie counterparts -- rely more heavily on $BL$ than on $RC$. Zombies seem to prefer debt with a maturity of two to five years, followed by short-term contracts of up to one year, which is most often associated with the intention to cover working capital needs \citep{AmbergJacobson2021}.

\begin{figure}[h!]
    %\vspace{-1cm}
    \caption{\normalsize{Share of Newly Granted Credit Sunk with Zombies -- Zombie-Definition $Z^{NAR}$}} \label{fig:CUS_FaceVal_Z_v3}
    \begin{center}
    \begin{subfigure}[t]{\textwidth}
    \centering
    \renewcommand\thesubfigure{\arabic{subfigure}}
    \begin{subfigure}[t]{0.45\textwidth}
        \centering
        \includegraphics[width=\textwidth, trim = 20mm 40mm 0mm 55mm, clip]{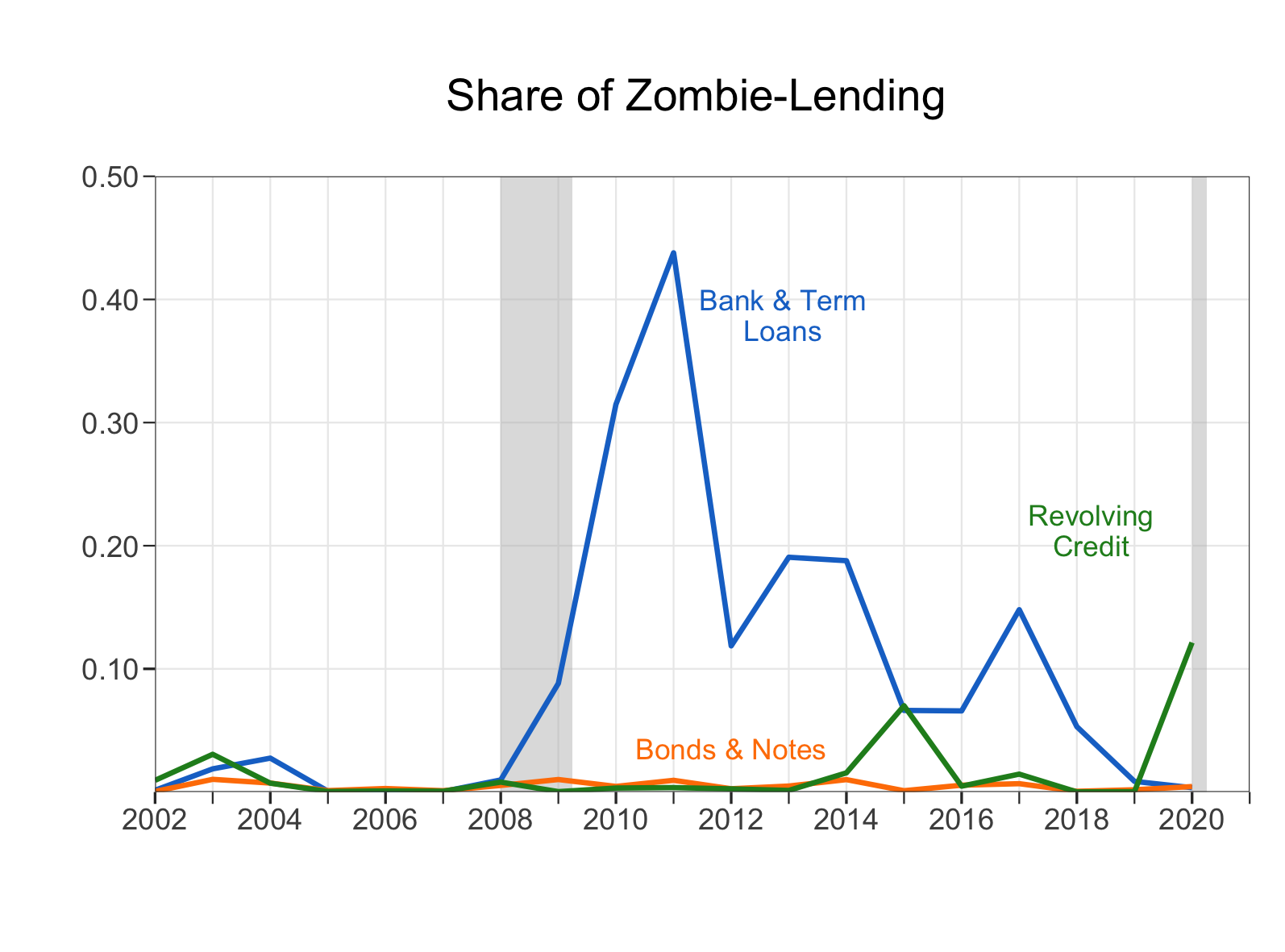}
        %\vspace{-0.1cm}
        \caption{1Q $\leq m \leq$ 4Q}
    \end{subfigure}%
    \begin{subfigure}[t]{0.45\textwidth}
        \centering
        \includegraphics[width=\textwidth, trim = 20mm 40mm 0mm 55mm, clip]{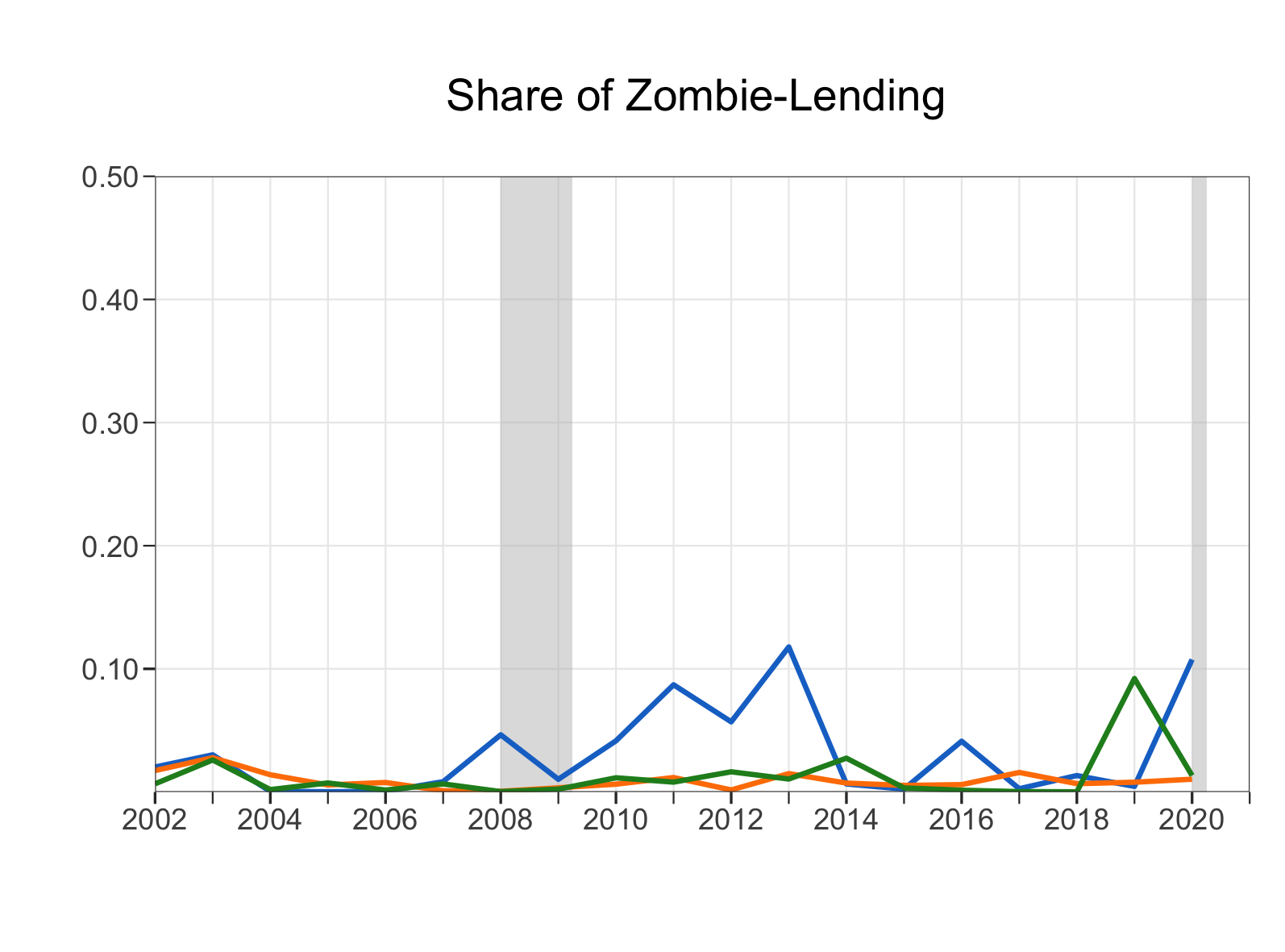}
        %\vspace{-0.1cm}
        \caption{5Q $\leq m \leq$ 8Q}
    \end{subfigure}
    \begin{subfigure}[t]{0.45\textwidth}
        \centering
        \includegraphics[width=\textwidth, trim = 20mm 40mm 0mm 55mm, clip]{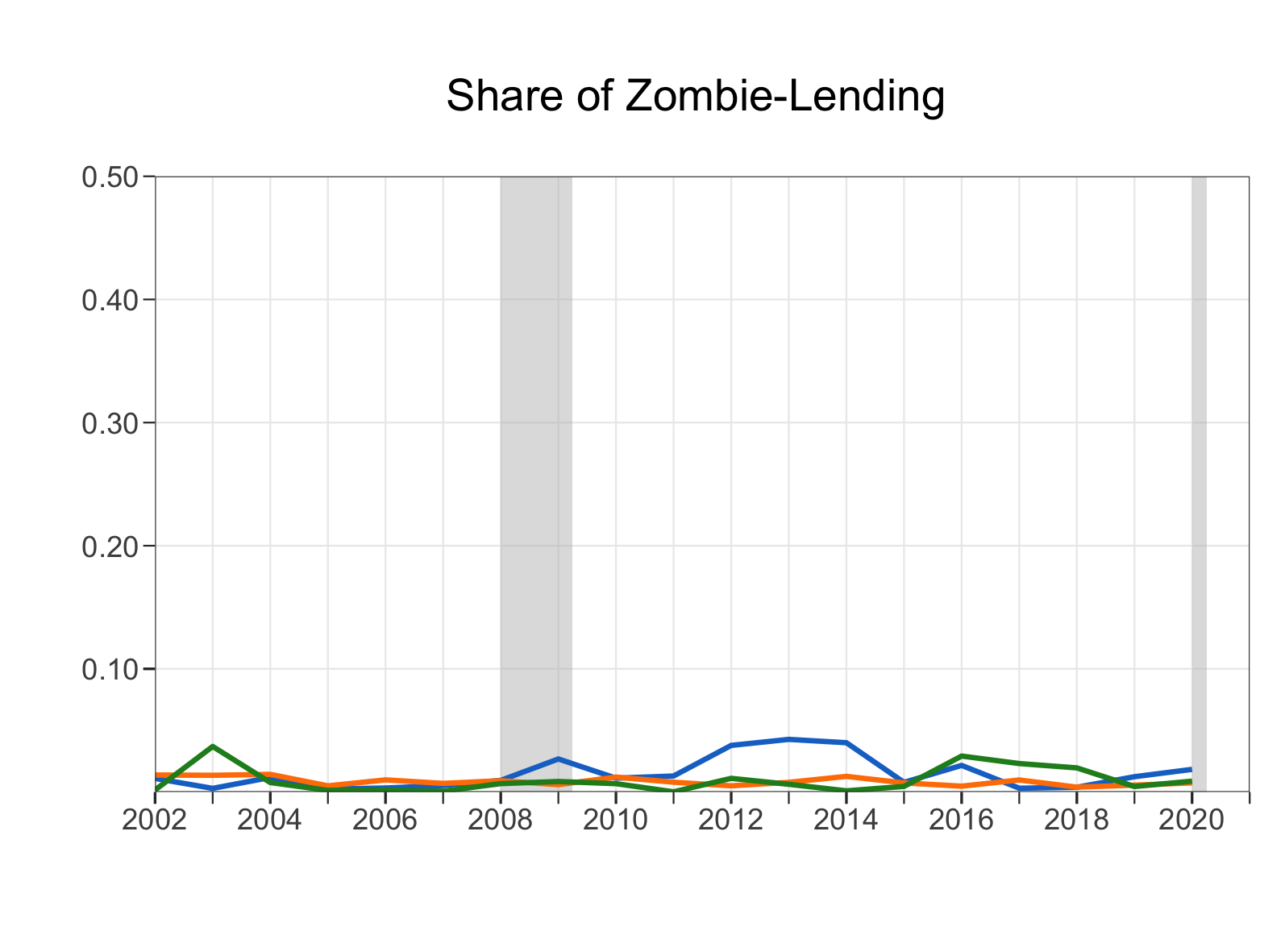}
        %\vspace{-0.1cm}
        \caption{9Q $\leq m \leq$ 20Q}
    \end{subfigure}%
    \begin{subfigure}[t]{0.45\textwidth}
        \centering
        \includegraphics[width=\textwidth, trim = 20mm 40mm 0mm 55mm, clip]{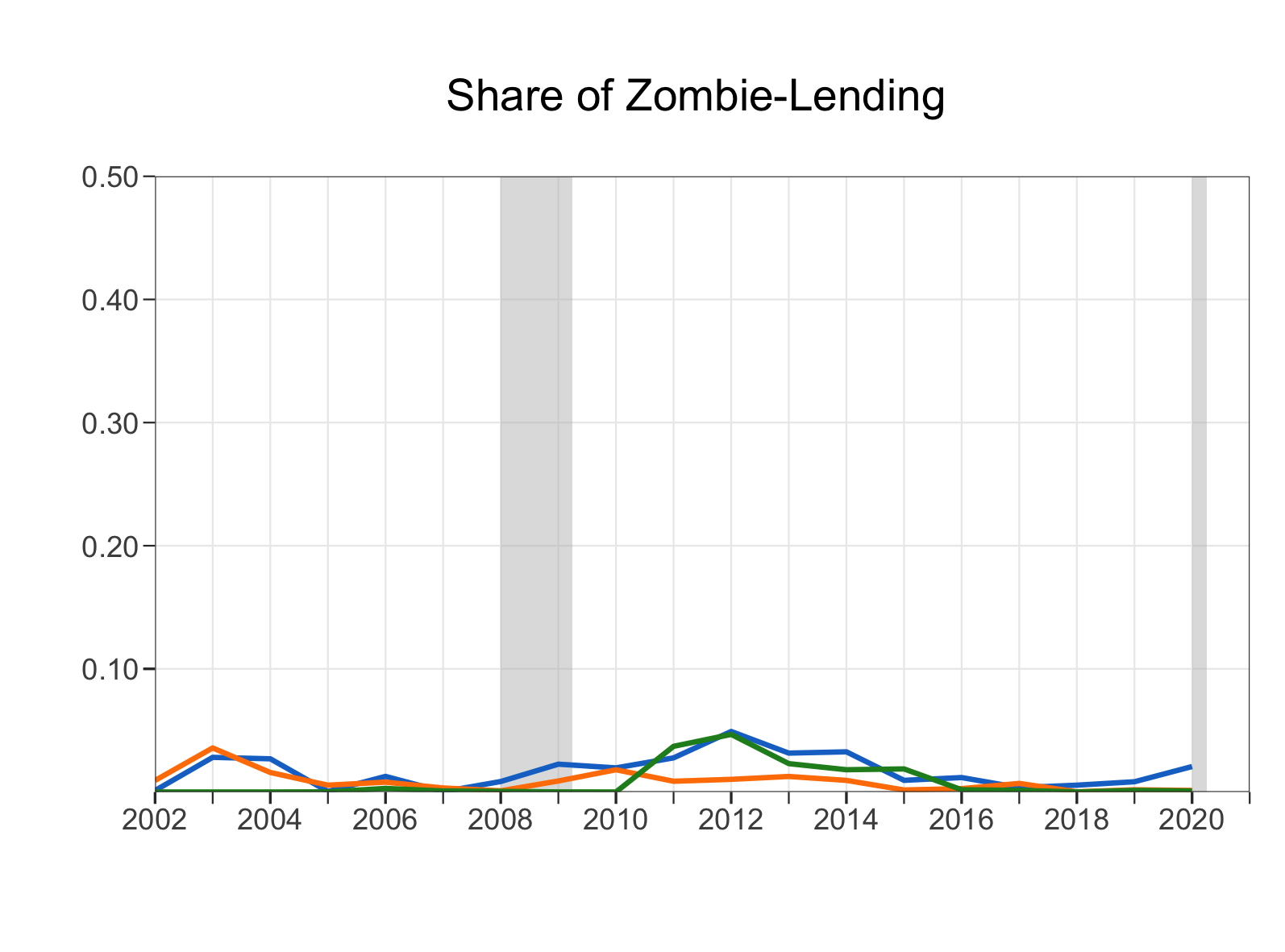}
        %\vspace{-0.1cm}
        \caption{21Q $\leq m \leq$ 40Q}
    \end{subfigure}
    \renewcommand\thesubfigure{\alph{subfigure}}
    \setcounter{subfigure}{0}
    \caption{\normalsize Full Sample}
    \end{subfigure}
    
    \vspace{0.5cm}
    \begin{subfigure}[t]{\textwidth}
    \centering
    \renewcommand\thesubfigure{\arabic{subfigure}}
    \setcounter{subfigure}{0}
    \begin{subfigure}[t]{0.45\textwidth}
        \centering
        \includegraphics[width=\textwidth, trim = 20mm 40mm 0mm 55mm, clip]{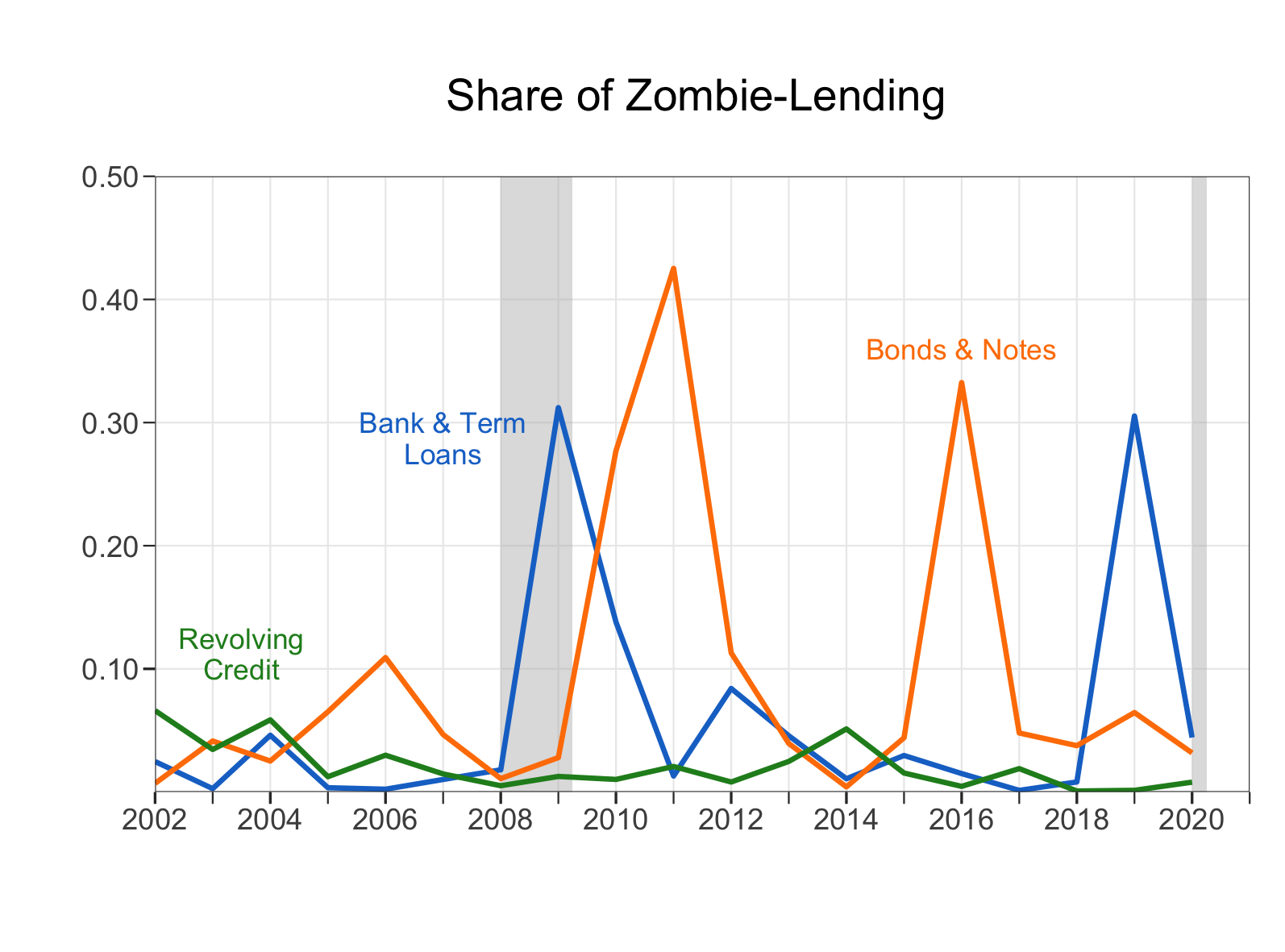}
        %\vspace{-0.1cm}
        \caption{1Q $\leq m \leq$ 4Q}
    \end{subfigure}%
    \begin{subfigure}[t]{0.45\textwidth}
        \centering
        \includegraphics[width=\textwidth, trim = 20mm 40mm 0mm 55mm, clip]{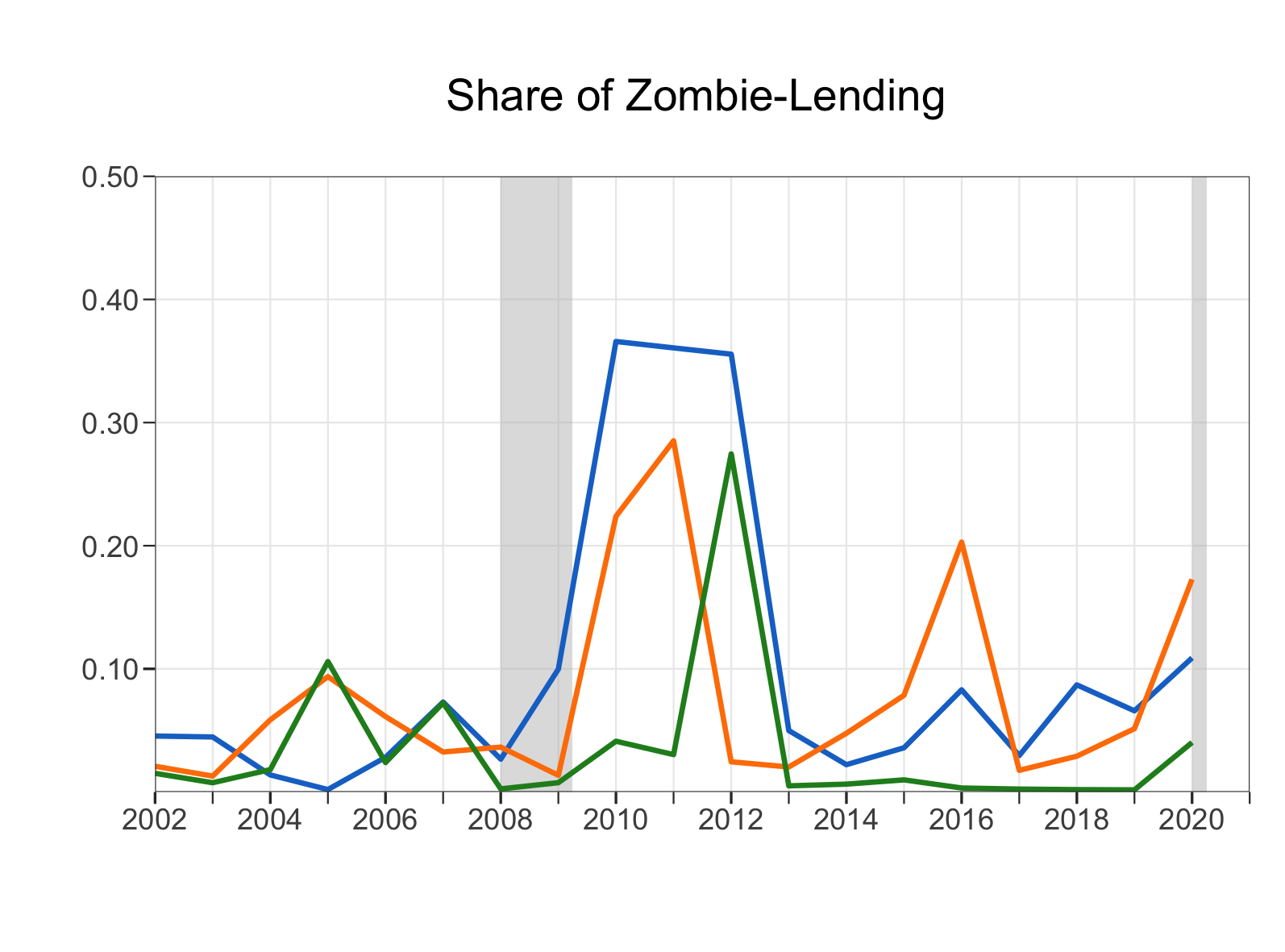}
        %\vspace{-0.1cm}
        \caption{5Q $\leq m \leq$ 8Q}
    \end{subfigure}
    \begin{subfigure}[t]{0.45\textwidth}
        \centering
        \includegraphics[width=\textwidth, trim = 20mm 40mm 0mm 55mm, clip]{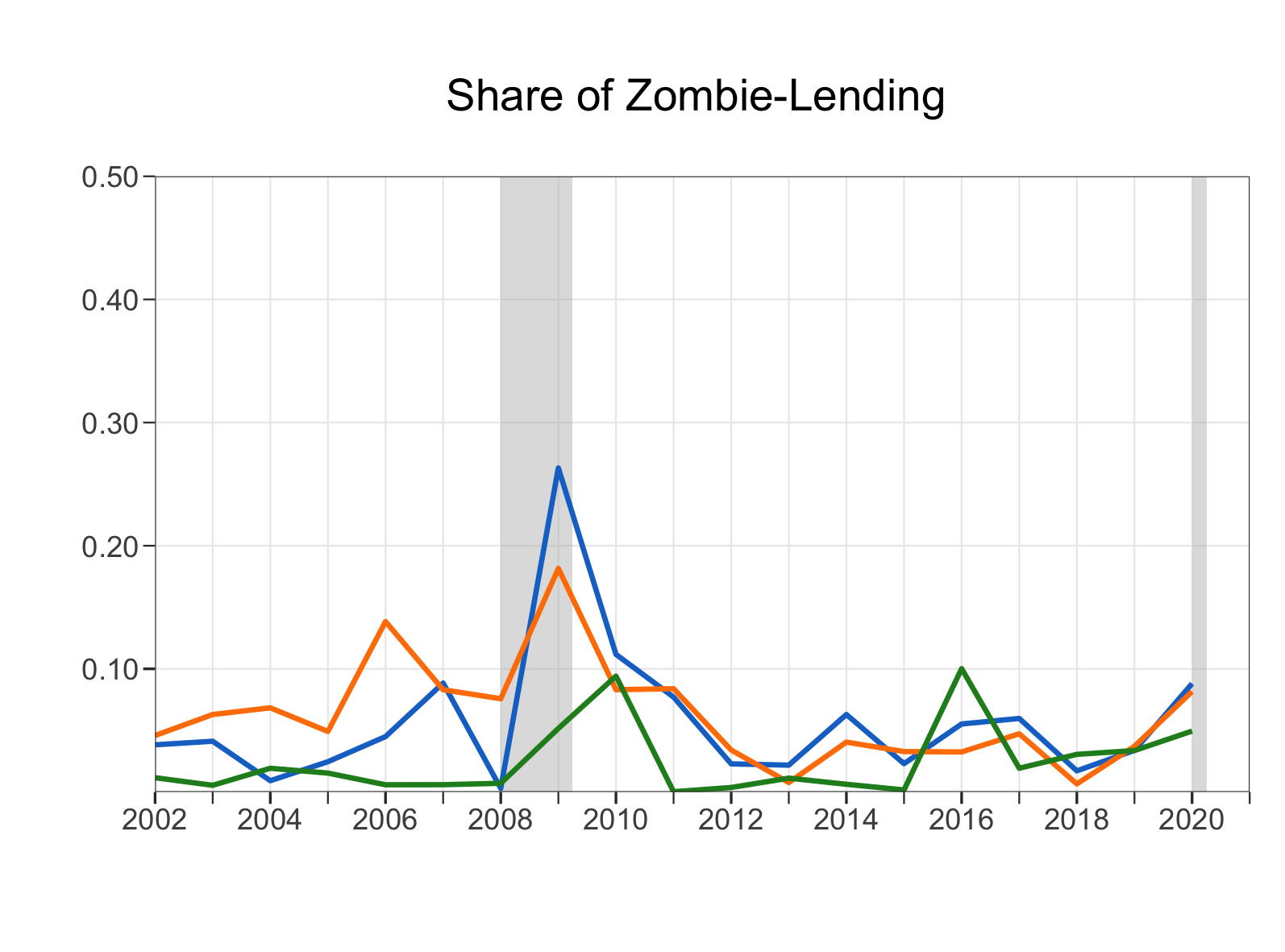}
        %\vspace{-0.1cm}
        \caption{9Q $\leq m \leq$ 20Q}
    \end{subfigure}%
    \begin{subfigure}[t]{0.45\textwidth}
        \centering
        \includegraphics[width=\textwidth, trim = 20mm 40mm 0mm 55mm, clip]{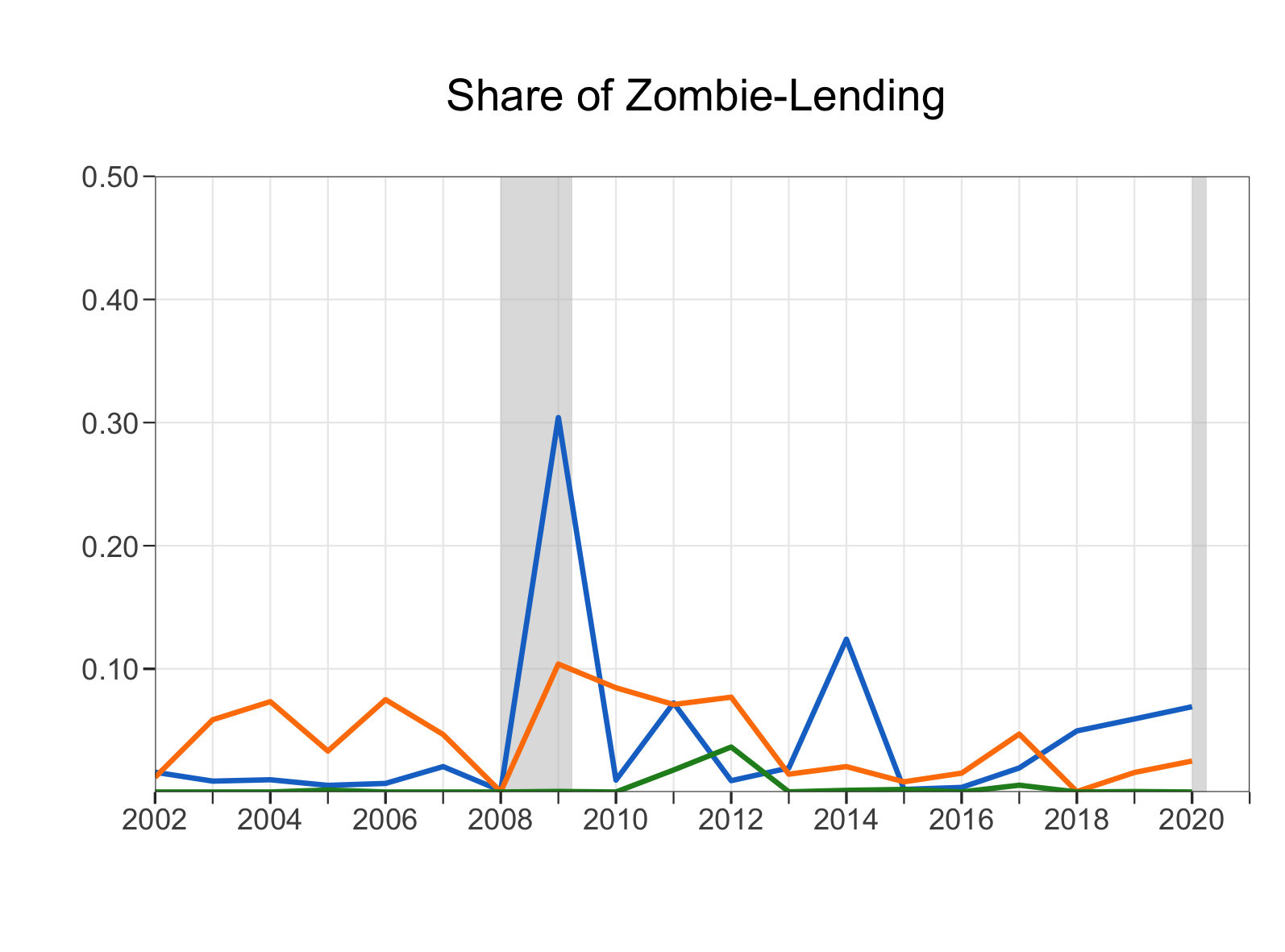}
        %\vspace{-0.1cm}
        \caption{21Q $\leq m \leq$ 40Q}
    \end{subfigure}
    \renewcommand\thesubfigure{\alph{subfigure}}
    \setcounter{subfigure}{1}
    \caption{\normalsize Small- \& Medium-Sized Companies}
    \end{subfigure}
    % \vspace*{0.5cm}
    % \begin{subfigure}[t]{0.5\textwidth}
    %     \centering
    %     \includegraphics[width=\textwidth, trim = 0mm 0mm 0mm 0mm, clip]{Figures/CUS_Lfraction_Z_legend.png}
    % \end{subfigure}
    
    \end{center}
    
\scriptsize{Source: Compustat and author's calculations. The shaded areas mark NBER recessions. We show the fraction of zombie-lending, i.e. the share of total first-time reported debt-obligations in year $t$, which were granted to zombie-firms. We show $BL$ in \textit{blue}, $RC$ in \textit{green}, and $BN$ in \textit{red}. }
\end{figure}

\noindent But is a major chunk of total credit sitting with zombie firms?   %Before turning to the empirical analysis, it is worth having a look at the share of total lending being granted to zombie firms. 
The upper panel in Figure \ref{fig:CUS_FaceVal_Z_v3} shows the share of total credit being granted to zombie firms for the three types of debt across different maturity buckets when applying the $Z^{NAR}$-definition. The graphs leave little ground for arguing that overall zombie-lending consumes a major part of the overall lending-pie. An observation that aligns with \cite{FavaraCameliaPOrive2022}.

\noindent Nevertheless, previous summary statistics have shown that zombies are rather small in size compared to the other companies in the sample. When it comes to lending, zombies may thus not necessarily compete with large but rather with small- and medium-sized companies.
The lower panel in Figure \ref{fig:CUS_FaceVal_Z_v3} shows the amount of zombie-credit relative to the total amount of debt granted to firms with less than 1,000 employees, which is our definition of small- and medium-sized companies (SMEs) inspired by \cite{ChodorowReich2013}. Within this size-category, zombie-lending is much more prevalent with shares rising to levels of up to 40\% of newly granted credit.

%Zombie-lending mounts up to over 40\% for short-term $BN$ in 2011, while the highest shares for $BL$ occur in the aftermath of the Great Recession with the share ranging around 35\% for contracts with a maturity falling into bucket (II). The share of new $RC$ granted to zombies is rarely exceeding 10\% in any maturity bucket, with the maximum share of almost 30\% being reported in 2012.

\noindent These simple descriptive results already provide valuable insights for the upcoming empirical analysis: judging on the grounds of the entire sample of Compustat firms, one may indeed form a prior on zombie-lending to not have any significant economic impact. For the class of SMEs, however, this perception may change.
In the following section, we will evaluate this supposition more formally.

%\clearpage
\section{Empirical Evidence on Zombie-Lending in the U.S.} \label{sec:Zombie_Evidence}

Over the past decade, advanced economies have experienced a period of low productivity growth. Recent research has linked this phenomenon to factors such as widening productivity dispersion across firms \citep{AndrewsEtAl2016}, rising capital misallocation \citep{GopinathEtAl2017}, subdued growth of non-zombies relative to zombies \citep{schivardi2017credit} and declining business dynamism \citep{DeckerEtAl2017}. 

\noindent Previous sections may have conveyed the impression that zombie prevalence to not be widespread among publicly listed U.S. companies. This does however not rule out the potential for zombification and zombie-lending to affect -- adversely or not -- the performance of non-zombies. In what follows, we test this hypothesis by targeting explicitly the root of zombie prevalence, namely the funding of non-viable firms. In particular, we assess the extent to which the share of fresh credit, granted to zombie-firms of a given industry $s$, affects productivity, capital-growth, and employment of their non-zombie peers. As described in Section \ref{sec:Data}, we allow for two types of credit: the first subsumes the sum of bank- and term-loans ($BL$) and revolving credit facilities ($RC$) under the notion of bank credit ($BC$); the second type captures credit taken up via capital markets in the form of Bonds \& Notes ($BN$). We further differentiate between short-term contracts with a maturity of up to four quarters, and long-term debt, i.e. contracts with a maturity of more than one year, but not exceeding ten years upon origination. As the upcoming sections will show, this differentiation helps to uncover the heterogeneity in the sensitivity of non-zombies' performance to the maturity structure of zombie-credit.

\noindent The remainder of this section is structured as follows: in Section \ref{sec:empirics_NZperformance} we assess the relationship between zombie-lending and its implications for productivity, and capital- and employment-growth of non-zombies. In Sections \ref{sec:empirics_prod}, \ref{sec:empirics_capgrowth}, and \ref{sec:empirics_emp}, we assess each performance measure more in-depth by splicing the group of non-zombies into more granular subgroups. Section \ref{sec:Z_ExitEntry} concludes our empirical analysis by inspecting the consequences of zombie-lending for business dynamism.

\subsection{Zombie-Lending and Non-Zombie Performance} 
\label{sec:empirics_NZperformance}
%We begin our empirical analysis by evaluating the spillovers of zombie-lending on the performance of non-zombies. 
As mentioned previously, we proxy ``performance'' by productivity, i.e. Total Factor Productivity (TFP), capital-growth, and employment-growth. We deploy the following models to formally describe these relationships for both short-term and longer-term zombie-credit:

\begin{align} \label{equ:NZperformance_Prod_Zlending}
    TFP_{i,t}  =  \mathbb{X} +  \beta_{BC} \; NZ_{i,t-1} \times BC^Z_{s,t-1}  +  \beta_{BN} \; NZ_{i,t-1} \times BN^Z_{s,t-1}   +  \varepsilon_{i,t}
\end{align}
\vspace{-1cm}
\begin{align} \label{equ:NZperformance_CapitalGrowth_Zlending}
    \Delta log\left(K_{i,t}\right)  =  \mathbb{X}  +  \beta_{BC} \; NZ_{i,t-1} \times BC^Z_{s,t-1}  +  \beta_{BN} \; NZ_{i,t-1} \times BN^Z_{s,t-1}  +  \varepsilon_{i,t}
\end{align}
\vspace{-1cm}
\begin{align} \label{equ:NZperformance_Employment_Zlending}
    \Delta EMP_{i,t}  =   \mathbb{X} +  \beta_{BC} \; NZ_{i,t-1} \times BC^Z_{s,t-1}  +  \beta_{BN} \;  NZ_{i,t-1} \times BN^Z_{s,t-1}  +  \varepsilon_{i,t}
\end{align}

\noindent As stated above, we use a firm's TFP ($TFP_{i,t}$) to analyze the relationship between zombie-lending and non-zombies' productivity. In Equation \eqref{equ:NZperformance_CapitalGrowth_Zlending} we measure the dependent variable ($\Delta log\left(K_{i,t}\right)$) as the first-difference of a firm's logged stock of (net) property, plant and equipment ($\Delta log\left(PPENT_{i,t}\right)$).
Lastly, we follow \cite{ChodorowReich2013} and measure employment-growth ($\Delta EMP_{i,t}$) by the symmetric growth formula (see Equation \eqref{equ:symm_growth} for further details).

\noindent All models contain a full set of controls ($\mathbb{X}$), which comprise firm-, sector-, year- and sector-year-fixed effects ($\alpha_i$ and $\delta_t$), a non-zombie indicator, $NZ_{i,t-1} \in \{0;1\}$, and a model-specific set of firm characteristics, which we describe in more detail in each of the corresponding subsections below.

\noindent Our main variable of interest is zombie-credit in the form of (i) $BC^Z_{s,t}$, which is the share of new bank-intermediated credit to industry $s$ that has been granted to non-viable firms in year $t$, and (ii) $BN^Z_{s,t}$, which is the share of new bonds and notes issued by zombie firms in industry $s$ in year $t$. These zombie-credit variables enter with a one period lag, in order to alleviate any endogeneity biases and to better motivate any causal relationships. Aggregating these lending variables to industry-shares is motivated by specifications found in the literature, which assess the effects zombification at the industry-level on the performance of healthy companies \citep{CaballeroHoshiKashyap2008,McGowanAndrewsMillot2018,schivardi2020identifying}. Furthermore, recent studies analyse the rise of specialized lenders, which grant credit to only a selected set of industries \citep{DiPattison2021}. Though this phenomenon is observed among small businesses, it aligns well with the specifications in upcoming sections. In order to directly assess the differential effect of zombie-lending on non-zombies' performance, the industry-shares of zombie-credit are interacted with an indicator for non-zombies ($NZ_{i,t-1}$). 
\noindent The results in Table \ref{tab:Model00_Z_v3} can be summarized fairly quickly: none of the models could establish a statistically significant relationship between zombie-lending and the performance of non-zombies -- neither for contracts with a maturity of up to four quarters upon origination, nor for longer-term debt. Overall, the results are grist to the mill of those questioning whether the observed adverse effects of zombification in Europe \citep{AndrewsEtAl2016,McGowanAndrews2018,schivardi2017credit,AcharyaEtAl2019} also hold in the case of the United States. Nevertheless, it might be worth digging a bit deeper.
%it might be premature to stop the investigation just here.
%take the short-cut and writing up the conclusion about spillover effects of zombie-lending on non-zombies' performance. 

\noindent Table \ref{tab:SumStat_Compustat_v3} documents stark differences in firm characteristics between zombies and viable companies in our sample. One of those is firm size -- whether measured in terms of total assets or employees. As asserted in Section \ref{sec:zombies_zombielending}, the class of viable firms may comprise companies, which do not directly compete with zombies in the search for appropriate funding. Lumping all non-zombies together, may therefore %suffer from insufficient granularity and 
overshadow the effects of zombie-lending for a certain type of non-zombie.
We will therefore continue our analysis by further splicing the group of non-zombies according to several firm characteristics, such as firm size, bank-dependency and access to bond markets.\footnote{See Figure \ref{fig:CUS_SMEs} for the share of these subgroups in our sample between 2002 and 2020.}
Other than in this subsection, we proceed by dedicating a single subsection to each of our three performance measures.

\begin{table}[H]
\renewcommand{\arraystretch}{0.75}
    \begin{center}
    %\vspace{-2cm}
    \captionsetup{justification=centering}
    \footnotesize
\caption{\label{tab:Model00_Z_v3} Regressions Results: Zombie-Lending \& Non-Zombie Performance \\ Zombie-Definition $Z^{NAR}$}

\begin{tabular}{l cccc} \hline
\addlinespace[2pt]
 Maturity ($m$) & & 1Q  $\leq m \leq$ 4Q & 1Q  $\leq m \leq$ 4Q 
 & 1Q  $\leq m \leq$ 4Q \\ 
 \cmidrule(lr){3-3} \cmidrule(lr){4-4} \cmidrule(lr){5-5}
 & & (1)  & (2) & (3)  \\
Variables & & $TFP_{i,t}$ & $\Delta log \left(K_{i,t}\right)$ & 
$\Delta EMP_{i,t}$  \\  \addlinespace[2pt] \hline

\addlinespace[10pt]

%%%%%%%%%%%%%%%%%%%%%%%%%%%%%%%%%%%%%%%%%%%
$ NZ_{i,t-1} $ & &
 -0.010 & 0.196*** & 0.780*** \\
%%%%%%%%%%%%%%%%%%%%%%%%%%%%%%%%%%%%%%%%%%%

\addlinespace[5pt]

%%%%%%%%%%%%%%%%%%%%%%%%%%%%%%%%%%%%%%%%%%%
$ NZ_{i,t-1} \times BC^Z_{s,t-1}$ & &
-0.100 & -0.257 & -0.227 \\ \addlinespace[5pt]

$ NZ_{i,t-1} \times BN^Z_{s,t-1} $& & 
0.202 & 0.091 & -0.227 \\ \addlinespace[5pt]
%%%%%%%%%%%%%%%%%%%%%%%%%%%%%%%%%%%%%%%%%%%

\addlinespace[10pt]

Years & & \multicolumn{3}{c}{2002 - 2020} \\ 
Observations & & 41,098 & 72,381 & 66,926 \\ 
Firms & & 5,826 & 9,709 & 8,928 \\   
Fixed Effects & & X &  X & X \\
Controls & & X &  X & X \\
Within-$R^2$ & & 0.03 & 0.12 & 0.05 \\   

\addlinespace[5pt]
    \midrule
\addlinespace[5pt]

Maturity ($m$) & & 5Q  $\leq m \leq$ 40Q & 5Q  $\leq m \leq$ 40Q 
 & 5Q  $\leq m \leq$ 40Q \\ 
 \cmidrule(lr){3-3} \cmidrule(lr){4-4} \cmidrule(lr){5-5}
 & & (4)  & (5) & (6)  \\
Variables & & $TFP_{i,t}$ & $\Delta log \left(K_{i,t}\right)$ & 
$\Delta EMP_{i,t}$  \\  \addlinespace[2pt] \hline

\addlinespace[10pt]

%%%%%%%%%%%%%%%%%%%%%%%%%%%%%%%%%%%%%%%%%%%
$ NZ_{i,t-1} $ & &
 -0.006 & 0.192*** & 0.081*** \\
%%%%%%%%%%%%%%%%%%%%%%%%%%%%%%%%%%%%%%%%%%%

\addlinespace[5pt]

%%%%%%%%%%%%%%%%%%%%%%%%%%%%%%%%%%%%%%%%%%%
$ NZ_{i,t-1} \times BC^Z_{s,t-1}$ & &
-0.476 & -0.407 & -0.007 \\ \addlinespace[5pt]

$ NZ_{i,t-1} \times BN^Z_{s,t-1} $& & 
-0.039 & 0.446 & -0.204 \\ \addlinespace[5pt]
%%%%%%%%%%%%%%%%%%%%%%%%%%%%%%%%%%%%%%%%%%%

\addlinespace[10pt]

Years & & \multicolumn{3}{c}{2002 - 2020} \\ 
Observations & & 41,100 & 72,466 & 67,006 \\ 
Firms & & 5,827 & 9,713 & 8,931 \\   
Fixed Effects & & X &  X & X \\
Controls & & X &  X & X \\
Within-$R^2$ & & 0.03 & 0.12 & 0.06 \\  

\hline
\addlinespace[1pt]
\end{tabular}
\end{center}
\noindent \tiny Notes: Each estimation includes firm-, industry-, year- and -industry-year-fixed effects. Standard errors are clustered at the firm-level.
% We restrict the sample to only those companies that reported either one of the bank-credit instruments, $BL$ or $RC$, or an issuance of \textit{bonds \& notes} (BN) at least once over the period 2002 - 2019. For these variables -- and those interacted with the non-zombie-indicator ($ BC^{Z}_{i,t}, BN^{Z}_{i,t} $). 
Controls are composed as follows: Models (1) and  (4) include a lagged measure of size ($log\left(AT_{i,t-1}\right)$) and a firm's lagged R\&D intensity ($XRD_{i,t-1} / AT_{i,t-1}$). 
Models (2) and  (5) include a lagged measure of size ($log\left(AT_{i,t-1}\right)$) and a firm's lagged asset tangibility ($PPENT_{i,t-1} / AT_{i,t-1}$)
Models (3) and  (6) include a lagged measure of size ($log\left(AT_{i,t-1}\right)$), a firm's lagged cash ratio ($CHE_{i,t-1} / LT_{i,t-1}$), it's profitability ($ROA_{i,t-1} = IB_{i,t-1} / AT_{i,t-1}$),and the lagged proxy for a firm's asset tangibility ($PPENT_{i,t-1} / AT_{i,t-1}$).
Robust standard errors in parentheses: *** p$<$0.01, ** p$<$0.05, * p$<$0.1\\
\end{table}

\subsubsection{Productivity} \label{sec:empirics_prod}
Our first in-depth analysis concerns the effects of zombie-lending on a firm's Total Factor Productivity ($TFP_{i,t}$). In order to divide the entire set of non-zombies into subgroups, we deploy the following model:

\vspace{-0.5cm}
\begin{align} \label{equ:Prod_Zlending}
    \begin{tabular}{lll}
    \( TFP_{i,t} = \) & 
    \(\beta_{BC} \; NZ_{i,t-1} \times D^\mathrm{t}_{i,t-1} \times D^\mathrm{i}_{i} \times BC^Z_{s,t-1} \) &
    \(+ \; \beta_X \; X_{i,t-1}\) \\ \addlinespace[5pt] 
    & \(+ \; \beta_{BN} \; NZ_{i,t-1} \times D^\mathrm{t}_{i,t-1} \times D^\mathrm{i}_{i} \times BN^Z_{s,t-1}\) 
    & \(+ \; \alpha_i \; + \; \delta_t \; + \; \varepsilon_{i,t}\)
    \end{tabular}
\end{align}

\noindent where $\alpha_i$ and $\delta_t$ comprise a full set of firm-, sector-, year- and sector-year-fixed effects. $NZ_{i,t-1} \in \{0;1\}$ is an indicator for firm $i$ being a viable firm, i.e. \textit{not} being classified as a zombie, in year $t$. 
The vector of firm-level covariates, $X_{i,t-1}$, comprises a measure of lagged company size ($log\left(AT_{i,t-1}\right)$) and a firm's lagged R\&D intensity ($XRD_{i,t-1}/AT_{i,t-1}$).

\noindent To recap, our main variables of interest are (i) the share of new $BC$ to industry $s$, which was granted to non-viable firms in year $t$, $BC^Z_{s,t-1}$, and (ii) the share of new bonds and notes, measured in U.S. dollars, that was issued by zombie firms in industry $s$ in year $t$ ($BN^Z_{s,t-1}$). These variables are interacted with several dummy variables. 
%The interaction with an indicator for non-zombies ($NZ_{i,t}$) captures the differential effect of zombie-lending on non-zombies' productivity. 
In addition to the indicator for non-zombies ($NZ_{i,t}$), we distinguish between two further sets of dummies: (i) the set of time-varying dummies, $D^\mathrm{t}_{i,t-1} = \{\mathbb{1}, SM_{i,t-1}\} $, where $SM_{i,t}$ symbolizes \textit{small-} \& \textit{medium-sized} firms (SME) following the definition in \cite{ChodorowReich2013}. Therein, ``small'' firms are defined as $EMP_{i,t} < 250$, and medium-sized firms as $250 \leq EMP_{i,t} < 1000$.; (ii) the set of firm-specific and time-invariant dummies ($D^\mathrm{i}_{i} = \{\mathbb{1}, bank.dep_{i}, bond.dep_{i}, bond_i, no.bond_i\} $, where $bank.dep_{i}$) indicates whether firm $i$ is bank-dependent, whereas $bond.dep_{i}$ flags those firms which primarily resort to the bond market for financing. Similarly, $bond_i$, $no.bond_i$ respectively, are indicators used to determine whether firm $i$ has access to the bond market. A firm is classified as bank-dependent, i.e. $bank.dep_{i} = 1$, if it relied more extensively on bank credit than on bonds and notes over the sample period, i.e.

\[
\text{if} \quad \sum^{2020}_{t=2002} BC_{i,t} > \sum^{2020}_{t=2002} BN_{i,t} \, \Rightarrow \, bank.dep_{i} = 1 \;.
\]

\noindent For $bond_i = 1$, firm $i$ is required to have reported the issuance of bonds -- conditional on the maturity bucket -- at least once throughout the sample period. Equation \eqref{equ:Prod_Zlending} collapses to Equation \eqref{equ:Prod_Zlending} in case of both dummies just being an identity-vector, i.e. each element $i$ being set to 1, such that $D^\mathrm{t}_{i,t-1} = D^\mathrm{i}_{i} = \mathbb{1}$ the first and third terms in Equation \ref{equ:Prod_Zlending} collapse to $NZ_{i,t-1} \times BC^Z_{s,t-1}$ and $NZ_{i,t-1} \times BN^Z_{s,t-1}$ respectively. %These are the least rigorous measures of zombie-lending, as they do not additionally discriminate between a firm's size, type of financial dependency, or financial market access.
Figure \ref{fig:CUS_SMEs} in Appendix \ref{app:CUS_SMEs} plots the share of SMEs and respective subgroups over the sample period 2002-2020.

\noindent The results are summarized in Table \ref{tab:TFP_Z_v3_L1}. Models (1)-(3) capture fresh inflows of short-term credit to non-viable firms and Models (4)-(6)) document the effects of longer-term funding. 
The first row in Table \ref{tab:TFP_Z_v3_L1} shows that, in general, viable firms are not more productive than their non-viable counterparts. Nevertheless, the negative spillovers from zombie-lending on $TFP$ of non-zombies can hardly be overlooked and occur in each of the six specifications.

\noindent In Model (1) we ask whether zombie-lending compromises productivity of non-zombie SMEs. The statistically significant relationship between $BN^Z_{s,t-1}$, interacted with the non-zombie and SME indicators, points to this specific subgroup seeing its productivity to drop by about $-0.513$ for every percentage point of an industry's previous period's share of new inflows in the form of short-term $BN$, that was allocated to zombie firms. 
To get a sense of the economic impact implied by these results, consider $BN^Z_{i,t-1}$ to increase by one standard deviation ($SD = 0.047$)\footnote{Standard-deviation is calculated over the sample period 2002-2020.}. This would cause $TFP$ of non-zombie SMEs to decrease by $SD \times \beta_{BN} \equiv 0.047 \times \left(-0.513\right) = -0.024$ -- relative to the control group. With the median $TFP$ of non-zombie SMEs being 0.157, this translates into a drop of -15.3\%.

\begin{landscape}
\begin{table}[h]
\renewcommand{\arraystretch}{0.75}
    %\begin{center}
    %\vspace{-2cm}
    \centering
     \begin{threeparttable}
    \captionsetup{justification=centering}
    \footnotesize
\caption{\label{tab:TFP_Z_v3_L1} Regressions Results: Total Factor Productivity \& Zombie-Lending \\ Zombie-Definition $Z^{NAR}$}

\begin{tabular}{l ccccccc} \hline
\addlinespace[2pt]
 Maturity ($m$) & & 1Q  $\leq m \leq$ 4Q & 1Q  $\leq m \leq$ 4Q 
 & 1Q  $\leq m \leq$ 4Q & 5Q  $\leq m \leq$ 40Q & 5Q  $\leq m \leq$ 40Q & 5Q  $\leq m \leq$ 40Q
  \\
 \cmidrule(lr){3-3} \cmidrule(lr){4-4} \cmidrule(lr){5-5}
 \cmidrule(lr){6-6} \cmidrule(lr){7-7} \cmidrule(lr){8-8}
 & & (1)  & (2) & (3) & (4) & (5) & (6) \\
Variables & & $TFP_{i,t}$ & $TFP_{i,t}$ & 
$TFP_{i,t}$ & $TFP_{i,t}$ & $TFP_{i,t}$ & $TFP_{i,t}$  \\  \addlinespace[2pt] \hline

\addlinespace[10pt]

%%%%%%%%%%%%%%%%%%%%%%%%%%%%%%%%%%%%%%%%%%%
$ NZ_{i,t-1} $ & &
0.011 & 0.057 & 0.032 & 0.014 & 0.036 & 0.045 \\
%%%%%%%%%%%%%%%%%%%%%%%%%%%%%%%%%%%%%%%%%%%

\addlinespace[5pt]

%%%%%%%%%%%%%%%%%%%%%%%%%%%%%%%%%%%%%%%%%%%
$ NZ_{i,t-1} \times SM_{i,t-1}$ & &
-0.019 & & & 0.000 & & \\ \addlinespace[5pt]

$ NZ_{i,t-1} \times SM_{i,t-1} \times BC^Z_{s,t-1}$ & &
0.072 & & & -0.987** & & \\ \addlinespace[5pt]

$ NZ_{i,t-1} \times SM_{i,t-1}  \times BN^Z_{s,t-1} $& & 
-0.513*** & & & -2.406*** & & \\ \addlinespace[5pt]
%%%%%%%%%%%%%%%%%%%%%%%%%%%%%%%%%%%%%%%%%%%

\addlinespace[5pt]

%%%%%%%%%%%%%%%%%%%%%%%%%%%%%%%%%%%%%%%%%%%
$ NZ_{i,t-1} \times SM_{i,t-1} \times bank.dep_{i}$ & &
 & -0.051 & & & 0.009 & \\ \addlinespace[5pt]
 
$ NZ_{i,t-1} \times SM_{i,t-1} \times bond.dep_{i}$ & &
 & 0.000 & & & -0.034 & \\ \addlinespace[5pt] 

$ NZ_{i,t-1} \times SM_{i,t-1} \times bank.dep_{i} \times BC^Z_{s,t-1}$ & &
 & -0.418 & & & -0.398 & \\ \addlinespace[5pt]
 
$ NZ_{i,t-1} \times SM_{i,t-1} \times bond.dep_{i} \times BN^Z_{s,t-1}$ & &
 & -0.871* & & & -4.647*** & \\ \addlinespace[5pt] 

%%%%%%%%%%%%%%%%%%%%%%%%%%%%%%%%%%%%%%%%%%%

\addlinespace[5pt]

%%%%%%%%%%%%%%%%%%%%%%%%%%%%%%%%%%%%%%%%%%%
$ NZ_{i,t-1} \times SM_{i,t-1} \times bank.dep_{i}  \times no.bond_{i}$ & &
 & & -0.020 & & & -0.031 \\ \addlinespace[5pt]

$ NZ_{i,t-1} \times SM_{i,t-1} \times bond.dep_{i}$ & & 
 & & 0.015 & & & -0.039 \\ \addlinespace[5pt]
 
$ NZ_{i,t-1} \times SM_{i,t-1} \times bank.dep_{i}  \times no.bond_{i} \times BC^Z_{s,t-1}$ & &
 & & -0.946** & & & 0.838 \\ \addlinespace[5pt]

$ NZ_{i,t-1} \times SM_{i,t-1} \times bond.dep_{i} \times BN^Z_{s,t-1}$ & & 
 & & -0.920* & & & -4.563*** \\ \addlinespace[5pt]
%%%%%%%%%%%%%%%%%%%%%%%%%%%%%%%%%%%%%%%%%%%

% %%%%%%%%%%%%%%%%%%%%%%%%%%%%%%%%%%%%%%%%%%%
% $ NZ_{i,t-1} \times \Delta log\left(BC^Z_{i,t}\right) \times SM_{i,t-1} \times bank.dep_{i}  \times no.bond_{i}$ & &
% & & & & & -- \\ \addlinespace[5pt]
 
% $ NZ_{i,t-1} \times \Delta log\left(BC^Z_{i,t}\right) \times SM_{i,t-1} \times bank.dep_{i}  \times bond_{i}$ & &
% & & & & & -- \\ \addlinespace[5pt]

% $ NZ_{i,t-1} \times \Delta log\left(BN^Z_{i,t}\right) \times SM_{i,t-1} \times bank.dep_{i} \times no.bond_{i}$ & & 
% & & & & & -- \\ \addlinespace[5pt]
 
% $ NZ_{i,t-1} \times \Delta log\left(BN^Z_{i,t}\right) \times SM_{i,t-1} \times bank.dep_{i} \times bond_{i}$ & & 
% & & & & & -- \\ \addlinespace[5pt]
% %%%%%%%%%%%%%%%%%%%%%%%%%%%%%%%%%%%%%%%%%%%

\addlinespace[10pt]

Years & & \multicolumn{6}{c}{2002 - 2020} \\ 
Observations & & 38,490 & 23,986 & 23,986 &  38,492 & 32,584 & 32,584 \\ 
Firms & & 5,396 & 2,727 & 2,727 &  5,397 & 4,081 & 4,081 \\   
Fixed Effects & & X &  X & X & X  & X & X \\
Controls & & X &  X & X & X  & X & X \\
Within-$R^2$ & & 0.03 & 0.04 & 0.04 &  0.03 & 0.03 & 0.03 \\ 

\hline
\addlinespace[1pt]
\end{tabular}
%\end{center}
\begin{tablenotes}
\tiny
\item[] Notes: Each estimation includes firm-, industry-, year- and -industry-year-fixed effects. Standard errors are clustered at the firm-level.
% We restrict the sample to only those companies that reported either one of the bank-credit instruments, $BL$ or $RC$, or an issuance of \textit{bonds \& notes} (BN) at least once over the period 2002 - 2019. For these variables -- and those interacted with the non-zombie-indicator ($ BC^{Z}_{i,t}, BN^{Z}_{i,t} $). 
Controls include a measure of lagged size ($log\left(AT_{i,t-1}\right)$) and a firm's lagged R\&D intensity ($XRD_{i,t-1} / AT_{i,t-1}$). Robust standard errors in parentheses: *** p$<$0.01, ** p$<$0.05, * p$<$0.1.
 \end{tablenotes}
 \end{threeparttable}
\end{table}
\end{landscape}

\noindent In Model (2) we further zoom in on bank-dependent ($bank.dep_i$) and bond-market dependent ($bond.dep_i$) non-zombie SMEs. Here, the negative spillovers from public debt markets' short-term zombie-lending activities increase by almost 70\% to $-0.881$. In contrast, from a statistical point of view, banks' engagement in zombie-credit activities does not signal any significant spillovers -- even when zooming in on bank-dependent, non-zombie SMEs.
This picture changes in Model (3): we restrict the set of bank-dependent, non-zombie SMEs even further by looking at the subgroup that has no access to bond markets. The median $TFP$ in that subgroup of highly bank-dependent viable firms is 0.128. Such a firm would see its productivity decrease by -40.6\%\footnote{See: $\frac{-0.946 \times 0.055}{0.128}\times 100 = -40.6$.} -- relative to the control group -- for a standard deviation (SD = 0.055) increase in an industry’s share of new credit inflows that is intermediated by banks and channeled to zombie firms.

\noindent Drawing comparisons on the $\beta_{BN}$ coefficients across Models (1) and (3) hints at the existence of another mechanism: 
the increasing magnitude of the point estimates suggests financial constraints amplifying the effect of the bond market's zombie-lending activities.
Models (4)-(6) reproduce the specifications of Models (1)-(3), but inspect the effects of longer-term contracts. The spillovers and amplification mechanism arising from the investment in bonds and notes issued by zombies are qualitatively similar to those observed in Models (1)-(3). 
Drawing a comparison in terms of economic impact, we conduct the same thought experiment as before, by looking at a one standard-deviation increase in newly granted industry-credit to zombies in the form of bonds and notes ($SD=0.026$) The coefficient on $\beta_{BN} = -4.563$ in Model (6) is attached to bond-market dependent non-zombies. The median $TFP$ of this subgroup is 0.145. Hence, a capital-market dependent non-zombie SME with median productivity would subsequently experience a decrease in $TFP$ of about -82\%\footnote{See: $\frac{-4.563 \times 0.026}{0.145}\times 100 = -81.1$.} for a one-standard deviation increase in $BN^Z_{s,t}$. Compared to Model (3), which looked at short-term zombie-lending, in which $TFP$ would decrease by roughly -30\%\footnote{See: $\frac{-0.920 \times 0.047}{0.145}\times 100 = -29.8$.}, the economic magnitude of the impact of longer-term zombie-credit is significantly more severe. 
For banks' engagement in zombie-lending, such an amplification channel is not present. Still, non-zombie SMEs are again not spared from negative spillovers.
%The fact that the statistical significant relationship between banks' zombie-lending activities, observed in Model (4), does not emerge in Models (5) and Model (6) suggests that the amplification mechanism via financial frictions, does not exist for zombie-debt with longer-term maturities. 

%\noindent Next, we turn to $TFP$ to also account for the productivity of the capital stock. This may give us a more comprehensive picture of the term \textit{productivity}, even though \cite{SargentRodriguez2000} argue that ``neither [labor productivity, TFP] should be relied upon exclusively''.

\noindent In a nutshell, Tables \ref{tab:Model00_Z_v3} and \ref{tab:TFP_Z_v3_L1} establish the following results: there exist statistically significant negative spillovers of zombie-lending on $TFP$, which however predominantly affect small- and medium-sized companies. Moreover, financial constraints amplify the economic impact of increased zombie-lending in the market of short-term bank-funding.
%with bank-dependent viable SMEs with no access to public debt markets being twice as sensitive to banks' zombie-lending activities than the group of bank-dependent viable SMEs as a whole. 
Lastly, the negative spillovers emerge primarily from the bond market, but do also arise via the banking-channel. The effects can both be found in short-term and longer-term debt contracts, with the economic impact.

\subsubsection{Capital Growth} \label{sec:empirics_capgrowth}
We now turn to the interaction between capital-growth and zombie-lending. 
Our model is designed along the lines of \cite{McGowanAndrewsMillot2018}, and collapses to Equation \eqref{equ:NZperformance_CapitalGrowth_Zlending} when setting each element $i$ in $D^\mathrm{t}_{i,t-1}$ and $D^\mathrm{i}_{i}$ equal to one:

\begin{align} \label{equ:CapitalGrowth_Zlending}
    \begin{tabular}{lll}
    \( \Delta log\left(K_{i,t}\right) = \) & 
    \(\beta_{BC} \; NZ_{i,t-1} \times D^\mathrm{t}_{i,t-1} \times D^\mathrm{i}_{i} \times BC^Z_{s,t-1} \) &
    \(+ \; \beta_X \; X_{i,t-1}\)
     \\ \addlinespace[5pt] 
    & \(+ \; \beta_{BN} \; NZ_{i,t-1} \times D^\mathrm{t}_{i,t-1} \times D^\mathrm{i}_{i} \times BN^Z_{s,t-1}\) 
    & \(+ \; \alpha_i \; + \; \delta_t \; + \; \varepsilon_{i,t}\)
    \end{tabular}
\end{align}

\noindent where $\alpha_i$ and $\delta_t$ again comprise a full set of firm-, sector-, year- and sector-year-fixed effects. %We follow \cite{GutierrezPhilippon2017} and compute the gross investment rate at the firm level in Compustat as follows: $\left(CAPX_{i,t} + XRD_{i,t}\right)/AT_{i,t}$. As in \cite{McGowanAndrewsMillot2018} we include a proxy for size and use the lagged log of assets ($log\left(AT_{i,t-1}\right)$), and lagged labor productivity, $LP_{i,t-1}$.
The dependent variable is capital-growth, measured as the first-difference of a firm's logged stock of (net) property, plant and equipment ($\Delta log\left(PPENT_{i,t}\right)$). The set of covariates accounts for lagged firm size and lagged asset tangibility. The time-varying ($D^\mathrm{t}_{i,t-1}$) and time-invariant ($D^\mathrm{i}_{i}$) dummies follow the description of Equation \eqref{equ:Prod_Zlending}.
As in the previous exercise on productivity, our main coefficients of interest are $\beta_{BC}$ and $\beta_{BN}$. We again lag the share of zombie-lending in industry $s$ by one period, in order to disentangle the strain of causality between reduced investment opportunities for non-zombies and an increase in the share of zombie-lending.

\noindent The negative effects in Table \ref{tab:PPENT_Z_v3} are not as dominant as those documented in Table \ref{tab:TFP_Z_v3_L1}.
Although non-zombies invest more than non-viable firms, there is no statistical ground for rejecting the hypothesis that short-term zombie-lending does not impair capital-growth of non-zombies. This might be explained by the fact that short-term debt is generally perceived to cover working-capital needs, whereas expenses for investment purposes are associated with longer-term maturities \citep{AmbergJacobson2021}. %\footnote{A rough text search of CapIQ's \textit{capitalstructuredescription} field, however, revealed that words such as "Equipment", "Capital Exp", and "Machine" were more frequently found among short-term contracts, while the longer-term contracts dominated for phrases such as "Real Estate" or "Mortgage".}. 
Models (4)-(6) are again only scarcely covered with statistically significant relationships.
Unlike for $TFP$, the negative spillovers build up exclusively in public debt markets and only affect non-zombie SMEs, which are capital-market dependent. The coefficient $\beta_{BN}$ in Models (5) and (6) quantifies investment of this specific subgroup of viable firms to decrease by about -1.85\%, for a one standard deviation increase in zombies' industry-share of newly issued longer-term bonds and notes.\footnote{For Model (5): $SD \times \beta_{BN} \equiv 0.026 \times \left(-0.718\right) \times 100 = -1.87\% $ and for Model (6) respectively: $SD \times \beta_{BN} \equiv 0.026 \times \left(-0.713\right) \times 100 = -1.85\%$.}

\noindent Though the negative consequences of zombie-lending are less pronounced when measuring non-zombies' performance in terms of capital-growth, a common denominator seems to build up: viable SMEs, which are reliant on funding provided by the bond market, are indeed sensitive to public debt markets' lending engagements to zombie firms.

\begin{landscape}
    \begin{table}[h!]
\renewcommand{\arraystretch}{0.75}
    %\begin{center}
    %\vspace{-2cm}
    \centering
     \begin{threeparttable}
    \captionsetup{justification=centering}
    \footnotesize
\caption{\label{tab:PPENT_Z_v3} Regressions Results: Capital Growth \& Zombie-Lending \\ Zombie-Definition $Z^{NAR}$}

\begin{tabular}{l ccccccc} \hline
\addlinespace[2pt]
 Maturity ($m$) & & 1Q  $\leq m \leq$ 4Q & 1Q  $\leq m \leq$ 4Q 
 & 1Q  $\leq m \leq$ 4Q & 5Q  $\leq m \leq$ 40Q & 5Q  $\leq m \leq$ 40Q & 5Q  $\leq m \leq$ 40Q
  \\
 \cmidrule(lr){3-3} \cmidrule(lr){4-4} \cmidrule(lr){5-5}
 \cmidrule(lr){6-6} \cmidrule(lr){7-7} \cmidrule(lr){8-8}
 & & (1)  & (2) & (3) & (4) & (5) & (6) \\
Variables & & $\Delta log \left(K_{i,t}\right)$ & $\Delta log \left(K_{i,t}\right)$ & 
$\Delta log \left(K_{i,t}\right)$ & $\Delta log \left(K_{i,t}\right)$ & $\Delta log \left(K_{i,t}\right)$ & $\Delta log \left(K_{i,t}\right)$  \\  \addlinespace[2pt] \hline

\addlinespace[10pt]

%%%%%%%%%%%%%%%%%%%%%%%%%%%%%%%%%%%%%%%%%%%
$ NZ_{i,t-1} $ & &
0.204*** & 0.175*** & 0.166*** & 0.205*** & 0.187*** & 0.175*** \\
%%%%%%%%%%%%%%%%%%%%%%%%%%%%%%%%%%%%%%%%%%%

\addlinespace[5pt]

%%%%%%%%%%%%%%%%%%%%%%%%%%%%%%%%%%%%%%%%%%%
$ NZ_{i,t-1} \times SM_{i,t-1}$ & &
-0.014 & & & -0.016 & & \\ \addlinespace[5pt]

$ NZ_{i,t-1} \times SM_{i,t-1} \times BC^Z_{s,t-1}$ & &
-0.027 & & & -0.225 & & \\ \addlinespace[5pt]

$ NZ_{i,t-1} \times SM_{i,t-1}  \times BN^Z_{s,t-1} $& & 
-0.180 & & & 0.256 & & \\ \addlinespace[5pt]
%%%%%%%%%%%%%%%%%%%%%%%%%%%%%%%%%%%%%%%%%%%

\addlinespace[5pt]

%%%%%%%%%%%%%%%%%%%%%%%%%%%%%%%%%%%%%%%%%%%
$ NZ_{i,t-1} \times SM_{i,t-1} \times bank.dep_{i}$ & &
 & -0.034 & & & -0.033* & \\ \addlinespace[5pt]
 
$ NZ_{i,t-1} \times SM_{i,t-1} \times bond.dep_{i}$ & &
 & -0.024 & & & 0.003 & \\ \addlinespace[5pt] 

$ NZ_{i,t-1} \times SM_{i,t-1} \times bank.dep_{i} \times BC^Z_{s,t-1}$ & &
 & 0.221 & & & 0.117 & \\ \addlinespace[5pt]
 
$ NZ_{i,t-1} \times SM_{i,t-1} \times bond.dep_{i} \times BN^Z_{s,t-1}$ & &
 & -0.222 & & & -0.718*** & \\ \addlinespace[5pt] 

%%%%%%%%%%%%%%%%%%%%%%%%%%%%%%%%%%%%%%%%%%%

\addlinespace[5pt]

%%%%%%%%%%%%%%%%%%%%%%%%%%%%%%%%%%%%%%%%%%%
$ NZ_{i,t-1} \times SM_{i,t-1} \times bank.dep_{i}  \times no.bond_{i}$ & &
 & & -0.036 & & & -0.032 \\ \addlinespace[5pt]

$ NZ_{i,t-1} \times SM_{i,t-1} \times bond.dep_{i}$ & & 
 & & -0.019 & & & 0.009 \\ \addlinespace[5pt]
 
$ NZ_{i,t-1} \times SM_{i,t-1} \times bank.dep_{i}  \times no.bond_{i} \times BC^Z_{s,t-1}$ & &
 & & 0.189 & & & 0.435 \\ \addlinespace[5pt]

$ NZ_{i,t-1} \times SM_{i,t-1} \times bond.dep_{i} \times BN^Z_{s,t-1}$ & & 
 & & -0.231 & & & -0.713** \\ \addlinespace[5pt]
%%%%%%%%%%%%%%%%%%%%%%%%%%%%%%%%%%%%%%%%%%%

% %%%%%%%%%%%%%%%%%%%%%%%%%%%%%%%%%%%%%%%%%%%
% $ NZ_{i,t-1} \times \Delta log\left(BC^Z_{i,t}\right) \times SM_{i,t-1} \times bank.dep_{i}  \times no.bond_{i}$ & &
% & & & & & -- \\ \addlinespace[5pt]
 
% $ NZ_{i,t-1} \times \Delta log\left(BC^Z_{i,t}\right) \times SM_{i,t-1} \times bank.dep_{i}  \times bond_{i}$ & &
% & & & & & -- \\ \addlinespace[5pt]

% $ NZ_{i,t-1} \times \Delta log\left(BN^Z_{i,t}\right) \times SM_{i,t-1} \times bank.dep_{i} \times no.bond_{i}$ & & 
% & & & & & -- \\ \addlinespace[5pt]
 
% $ NZ_{i,t-1} \times \Delta log\left(BN^Z_{i,t}\right) \times SM_{i,t-1} \times bank.dep_{i} \times bond_{i}$ & & 
% & & & & & -- \\ \addlinespace[5pt]
% %%%%%%%%%%%%%%%%%%%%%%%%%%%%%%%%%%%%%%%%%%%

\addlinespace[10pt]

Years & & \multicolumn{6}{c}{2002 - 2020} \\ 
Observations & & 66,133 & 42,062 & 42,062 & 66,216 & 56,644 & 56,644 \\ 
Firms & & 8,792 & 4,474 & 4,474 &  8,796 & 6,647 & 6,647 \\   
Fixed Effects & & X &  X & X & X  & X & X \\
Controls & & X &  X & X & X  & X & X \\
Within-$R^2$ & & 0.11 & 0.12 & 0.12 &  0.12 & 0.12 & 0.12 \\

\hline
\addlinespace[1pt]
\end{tabular}
%\end{center}\\
 \begin{tablenotes}
\tiny
\item[]  Notes: Each estimation includes firm-, industry-, year- and -industry-year-fixed effects. Standard errors are clustered at the firm-level.
%We restrict the sample to only those companies that reported either one of the bank-credit instruments, $BL$ or $RC$, or an issuance of \textit{bonds \& notes} (BN) at least once over the period 2002 - 2019. For these variables -- and those interacted with the non-zombie-indicator ($ BC^{Z}_{i,t}, BN^{Z}_{i,t} $). 
Controls include a lagged measure of size ($log\left(AT_{i,t-1}\right)$) and a firm's lagged asset tangibility ($PPENT_{i,t-1} / AT_{i,t-1}$).  Robust standard errors in parentheses:  *** p$<$0.01, ** p$<$0.05, * p$<$0.1.
 \end{tablenotes}
 \end{threeparttable}
\end{table}
\end{landscape}

\subsubsection{Employment} \label{sec:empirics_emp}
Lastly, we evaluate the effects of zombie-lending on non-zombies' employment-growth. \cite{ChodorowReich2013} finds that changes in credit supply during the Great Recession materialized in diminished employment-growth, with SMEs being most affected. Moreover, having no access to bond markets aggravated the role of banks' credit supply in firms' employment-growth.
The findings in Sections \ref{sec:empirics_prod} and \ref{sec:empirics_capgrowth} suggested that zombie-lending impairs the performance of non-zombies via similar transmission channels.
To test whether this observation also holds in terms of non-zombies' employment-growth, we deploy the following model\footnote{As in the previous sections, this equation collapses to Equation \eqref{equ:NZperformance_Employment_Zlending} when setting each element $i$ in $D^\mathrm{t}_{i,t-1}$ and $D^\mathrm{i}_{i}$ equal to one.}: 
%\vspace{-0.15cm}
\begin{align} \label{equ:Employment_Zlending}
\begin{tabular}{lll}
    \(\Delta EMP_{i,t} = \) & 
    \(\beta_{BC} \; NZ_{i,t-1} \times D^\mathrm{t}_{i,t-1} \times D^\mathrm{i}_{i} \times BC^Z_{s,t-1} \) &
    \(+ \; \beta_X \; X_{i,t-1}\)
     \\ \addlinespace[5pt] 
    & \(+ \; \beta_{BN} \; NZ_{i,t-1} \times D^\mathrm{t}_{i,t-1} \times D^\mathrm{i}_{i} \times BN^Z_{s,t-1}\) 
    & \(+ \; \alpha_i \; + \; \delta_t \; + \; \varepsilon_{i,t}\)
\end{tabular}
\end{align}

\noindent where $\alpha_i$ and $\delta_t$ again comprise a full set of firm-, sector-, year- and sector-year-fixed effects. The set of controls, $X_{i,t-1}$, includes a lagged indicator for non-zombies, $NZ_{i,t-1}$, and several firm characteristics such as a lagged measure of size ($log\left(AT_{i,t-1}\right)$), a firm's lagged cash ratio ($CHE_{i,t-1} / LT_{i,t-1}$), a firm's ROA as a proxy for profitability ($IB_{i,t-1} / AT_{i,t-1}$), and the lagged measure for a firm's asset tangibility ($PPENT_{i,t-1} / AT_{i,t-1}$). The dependent variable $\Delta EMP_{i,t}$ is our measure of employment growth, measured by the symmetric growth formula as in \cite{ChodorowReich2013}:
%\vspace{-0.15cm}
\begin{align} \label{equ:symm_growth}
    \Delta EMP_{i,t} = \frac{EMP_{i,t} - EMP_{i,t-1}}{0.5 \times \left(EMP_{i,t} + EMP_{i,t-1}\right)}
\end{align}

\noindent The results in Table \ref{tab:EMPgr_Z_v3} carry over the impression from Table \ref{tab:PPENT_Z_v3}: there are negative spillovers from zombie-lending on non-zombies, however, the effects are not as profuse as they are for $TFP$. The only statistically significant relationship between zombie-lending and non-zombies employment-growth is only emerging in Model (6), i.e. for the subgroup of bank-dependent, non-zombie SMEs, which do not access the bond-market.
Similar to capital-growth, the effects can only be found among debt with longer-term maturities. Yet, the spillovers do not arise from the bond-market, but are emerging from banks' engagement in zombie-lending.

\noindent Quantifying the impact of zombie-lending on employment-growth in economic terms, the $\beta_{BC}$ estimates in Model (6) suggest employment growth to decrease by about -0.82\% for a one standard deviation increase in $BC^Z_{s,t}$.\footnote{See: $SD \times \beta_{BC} \equiv 0.025 \times \left(-0.326\right) \times 100 = -0.82\% $}
Even if $\beta_{BC}$ is not statistically significant in Models (4) and (5), we can again observe the amplification mechanism, caused by financial frictions, at play: the subgroup of bank-dependent, non-zombie SMEs, which does not enjoy access to the bond market, is most severely impacted by zombie-lending activities of the banking sector. 

\begin{landscape}
\begin{table}[h!]
\renewcommand{\arraystretch}{0.75}
    \centering
     \begin{threeparttable}
    \captionsetup{justification=centering}
    \footnotesize
\caption{\label{tab:EMPgr_Z_v3} Regressions Results: Employment Growth \& Zombie-Lending \\ Zombie-Definition $Z^{NAR}$}

\begin{tabular}{l ccccccc} \hline
\addlinespace[2pt]
 Maturity ($m$) & & 1Q  $\leq m \leq$ 4Q & 1Q  $\leq m \leq$ 4Q 
 & 1Q  $\leq m \leq$ 4Q & 5Q  $\leq m \leq$ 40Q & 5Q  $\leq m \leq$ 40Q & 5Q  $\leq m \leq$ 40Q
  \\
 \cmidrule(lr){3-3} \cmidrule(lr){4-4} \cmidrule(lr){5-5}
 \cmidrule(lr){6-6} \cmidrule(lr){7-7} \cmidrule(lr){8-8}
 & & (1)  & (2) & (3) & (4) & (5) & (6) \\
Variables & & $\Delta EMP_{i,t}$ & $\Delta EMP_{i,t}$ & $\Delta EMP_{i,t}$ & 
$\Delta EMP_{i,t}$ & $\Delta EMP_{i,t}$ & $\Delta EMP_{i,t}$  \\  \addlinespace[2pt] \hline

\addlinespace[10pt]

%%%%%%%%%%%%%%%%%%%%%%%%%%%%%%%%%%%%%%%%%%%
$ NZ_{i,t-1} $ & &
0.009 & 0.010 & 0.040*** & 0.010 & 0.016 & 0.046*** \\
%%%%%%%%%%%%%%%%%%%%%%%%%%%%%%%%%%%%%%%%%%%

\addlinespace[5pt]

%%%%%%%%%%%%%%%%%%%%%%%%%%%%%%%%%%%%%%%%%%%
$ NZ_{i,t-1} \times SM_{i,t-1}$ & &
0.083*** & & & 0.082*** & & \\ \addlinespace[5pt]

$ NZ_{i,t-1} \times SM_{i,t-1} \times BC^Z_{s,t-1}$ & &
0.019 & & & -0.053 & & \\ \addlinespace[5pt]

$ NZ_{i,t-1} \times SM_{i,t-1}  \times BN^Z_{s,t-1} $& & 
-0.034 & & & 0.026 & & \\ \addlinespace[5pt]
%%%%%%%%%%%%%%%%%%%%%%%%%%%%%%%%%%%%%%%%%%%

\addlinespace[5pt]

%%%%%%%%%%%%%%%%%%%%%%%%%%%%%%%%%%%%%%%%%%%
$ NZ_{i,t-1} \times SM_{i,t-1} \times bank.dep_{i}$ & &
 & 0.093*** & & & 0.077*** & \\ \addlinespace[5pt]
 
$ NZ_{i,t-1} \times SM_{i,t-1} \times bond.dep_{i}$ & &
 & 0.081*** & & & 0.079*** & \\ \addlinespace[5pt] 

$ NZ_{i,t-1} \times SM_{i,t-1} \times bank.dep_{i} \times BC^Z_{s,t-1}$ & &
 & 0.0278 & & & -0.197 & \\ \addlinespace[5pt]
 
$ NZ_{i,t-1} \times SM_{i,t-1} \times bond.dep_{i} \times BN^Z_{s,t-1}$ & &
 & -0.076 & & & 0.282 & \\ \addlinespace[5pt] 

%%%%%%%%%%%%%%%%%%%%%%%%%%%%%%%%%%%%%%%%%%%

\addlinespace[5pt]

%%%%%%%%%%%%%%%%%%%%%%%%%%%%%%%%%%%%%%%%%%%
$ NZ_{i,t-1} \times SM_{i,t-1} \times bank.dep_{i}  \times no.bond_{i}$ & &
 & & 0.082*** & & & 0.049*** \\ \addlinespace[5pt]

$ NZ_{i,t-1} \times SM_{i,t-1} \times bond.dep_{i}$ & & 
 & & 0.065*** & & & 0.064*** \\ \addlinespace[5pt]
 
$ NZ_{i,t-1} \times SM_{i,t-1} \times bank.dep_{i}  \times no.bond_{i} \times BC^Z_{s,t-1}$ & &
 & & -0.015 & & & -0.326** \\ \addlinespace[5pt]

$ NZ_{i,t-1} \times SM_{i,t-1} \times bond.dep_{i} \times BN^Z_{s,t-1}$ & & 
 & & -0.081 & & & 0.277 \\ \addlinespace[5pt]
%%%%%%%%%%%%%%%%%%%%%%%%%%%%%%%%%%%%%%%%%%%

\addlinespace[10pt]

Years & & \multicolumn{6}{c}{2002 - 2020} \\ 
Observations & & 66,926 & 42,360 & 42,360 &  67,006 & 56,922 & 56,922 \\ 
Firms & & 8,928 & 4,494 & 4,494 &  8,931 & 6,680 & 6,680 \\   
Fixed Effects & & X &  X & X & X  & X & X \\
Controls & & X &  X & X & X  & X & X \\
Within-$R^2$ & & 0.06 & 0.07 & 0.07 &  0.06 & 0.07 & 0.07 \\ 

\hline
\addlinespace[1pt]
\end{tabular}
\noindent \tiny Notes: Each estimation includes firm-, industry-, year- and -industry-year-fixed effects. Standard errors are clustered at the firm-level.
%We restrict the sample to only those companies that reported either one of the bank-credit instruments, $BL$ or $RC$, or an issuance of \textit{bonds \& notes} (BN) at least once over the period 2002 - 2019. For these variables -- and those interacted with the non-zombie-indicator ($ BC^{Z}_{i,t}, BN^{Z}_{i,t} $). 
Controls include a lagged measure of size ($log\left(AT_{i,t-1}\right)$), a firm's lagged cash ratio ($CHE_{i,t-1} / LT_{i,t-1}$), it's profitability ($ROA_{i,t-1} = IB_{i,t-1} / AT_{i,t-1}$),and the lagged proxy for a firm's asset tangibility ($PPENT_{i,t-1} / AT_{i,t-1}$). Robust standard errors in parentheses: *** p$<$0.01, ** p$<$0.05, * p$<$0.1 \\
 \end{threeparttable}
\end{table}
\end{landscape}

\noindent Before moving on to the last part of our empirical analysis, we briefly summarize the findings so far. The applications of this section have revealed statistically and economically significant spillovers of zombie-lending on the performance of non-zombies. Yet, the effects -- found under our specification -- are most prominent for $TFP$ and are only marginally emerging for investment and employment-growth.
Still, this leads us to conclude that zombification is a phenomenon that is not to be dismissed as non-existent in the U.S. economy.  
It is, however, primarily small- and medium-sized healthy companies that are exposed to the negative spillovers of granting credit to non-viable firms. Furthermore, different metrics of performance respond to different forms of zombie-lending: $TFP$ is susceptible to zombie-lending activities by both banks and capital markets by means of short- and longer-term debt. In contrast, capital-growth is only sensitive to zombie-lending in the form of longer-term financing through bonds and notes. Short-term financing and banks' engagement in zombie-lending does not seem to matter from a statistical point of view. Regarding employment-growth, the negative spillovers emerge exclusively via the bank-lending channel in the form of again longer-term debt contracts.
Lastly, the results also reveal the existence of financial frictions amplifying the negative effects of short-term zombie-lending. %, with the subgroups of bank-dependent SMEs, and especially those with no access to the bond market, being most severely affected.
Even though Figures \ref{fig:CUS_FaceVal_Z_v3} and \ref{fig:ZvsZ3} may have conveyed the impression of zombie prevalence not being a defining characteristic of the U.S. economy, the results of this subsection suggest that the performance of certain subgroups of corporations does indeed suffer from the negative spillovers of zombie-lending activities.

\noindent In the next section, we examine whether zombie-lending also has its saying in the functioning of another mechanism, whose demise is said to have contributed to the sluggish productivity-growth over the past decade: business dynamism. 

\subsection{Zombie-Lending and Business Dynamism} \label{sec:Z_ExitEntry}
Business dynamism describes the replacement of non-viable firms by new and striving entrants. This cleansing effect is supposed to serve as a catalyst of economic growth. Its slowing down is regarded as one reason for weak productivity-growth in recent decades. One explanation for this downward trend is the inefficient allocation of available resources \citep{DeckerEtAl2017}. , 
%by replacing unproductive and unprofitable companies with new thriving entrants. However, 
Keeping certain firms artificially alive, i.e. by granting them new funding, congests this growth engine: the prevalence of zombies diminishes the availability of capital and labor in the economy, thereby putting downward pressure on firms' profits and increasing the threshold for new firms to enter the market \citep{CaballeroHoshiKashyap2008}.  

\begin{figure}[h!]
    \begin{center}
    \caption{\footnotesize{Zombies vs. Newbies -- Zombie-Definition $Z^{NAR}$}} \label{fig:Z3vsNB}
        \includegraphics[width=0.75\textwidth, trim = 0mm 60mm 0mm 65mm, clip]{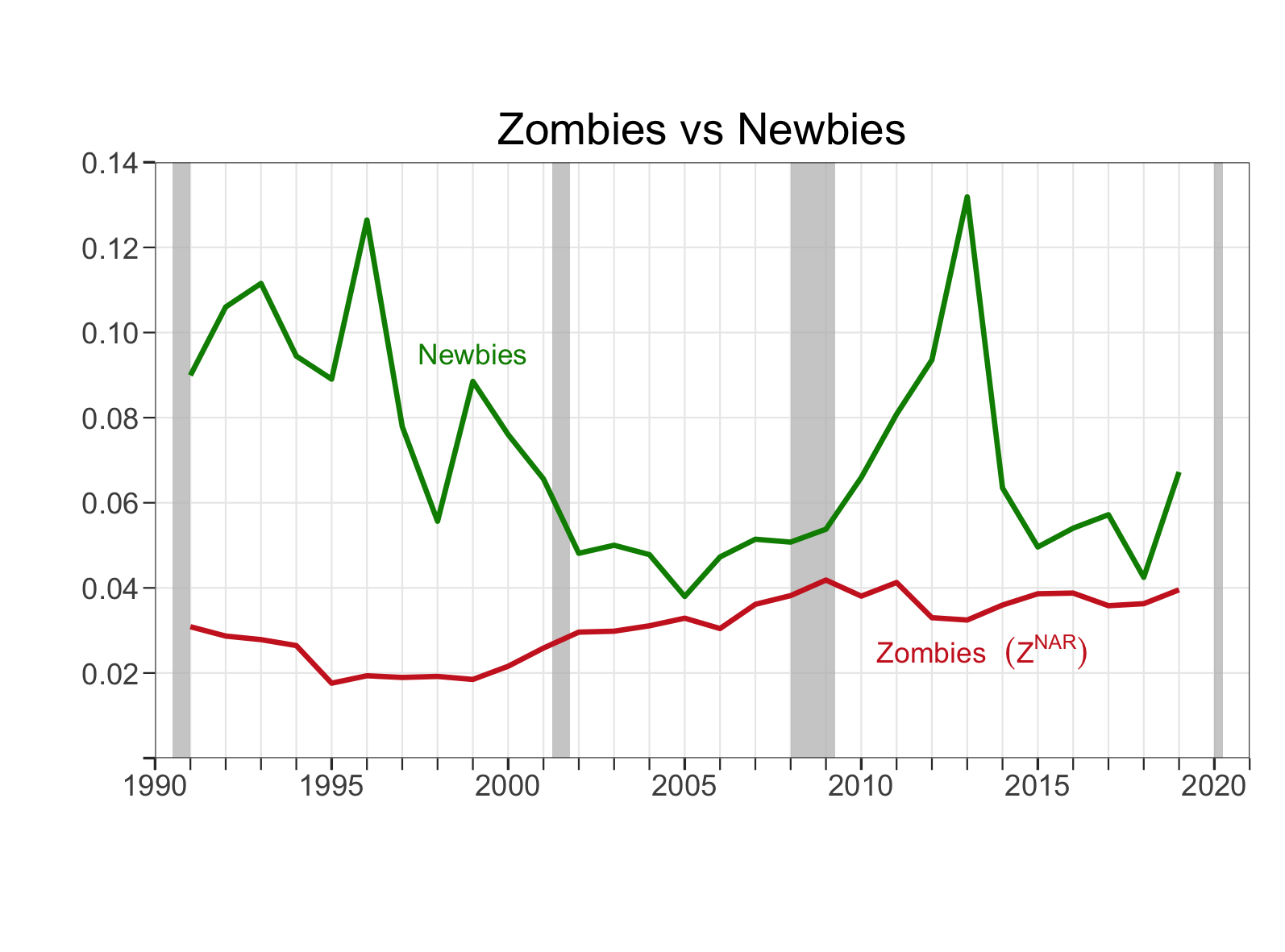}
        \end{center}
\begin{fignote} 
	\centering \hspace{1cm}
\begin{minipage}{.65\textwidth}
	\scriptsize Notes: Shaded areas mark NBER recessions. The denominator for both series includes all firms in a given year, whose data reporting allows for an assessment of their zombie-status. That is, the total number of individual firms in a given year may be larger than the set of firms, for which a statement about their zombie-status is possible, due to missing data.
\end{minipage}
\end{fignote}

\end{figure}

\noindent Figure \ref{fig:Z3vsNB} shows the share of zombies and new entrants (\textit{newbies}, $NB_t$) in each year between 1991 and 2018. While the share of zombies increased steadily over the course of the mid 1990s until the aftermath of the Great Recession, the share of newbies per year fell sharply starting at the end of the 1990s, recovered in the years after the financial crisis, before retreating again. 
The movements of the newbies- and zombie-share in the first half of that period corroborates the slowing down of business dynamism with both shares gradually converging. Even though the fraction of zombies -- under our definition of $Z^{NAR}$ -- never surpasses the share of newbies, the difference between the two has started to narrow down again from 2015 onward. 

\noindent We therefore continue our empirical analysis to understand whether zombie-lending affects entry decisions of new firms. To do so, we aggregate the firm-level observations to two-digits NAICS industry-level data and set up the following model:

\begin{equation} \label{equ:cleansing}
    \begin{tabular}{lll}
    \(\frac{NB_{s,t}}{N_{s,t}} = \) &
    \(\beta_{BC} \; BC^Z_{s,t-1} \) &
    \(+ \; \beta_{BN} \; BN^Z_{s,t-1} \) \\ \addlinespace[2pt]
    & \(+ \; \beta_{BC^B} \; bank.dep_{s,t-1} \times BC^Z_{s,t-1} \) &
    \(+ \; \beta_{BN^{CM}} \; bond.dep_{s,t-1} \times BN^Z_{s,t-1} \) \\ \addlinespace[2pt]
    & \(+ \; \beta_{NB} \times \frac{NB_{s,t-1}}{N_{s,t-1}} \) &
    \(+ \;\alpha_s \; + \; \delta_t \; + \; \varepsilon_{s,t} \; , \)
    \end{tabular}
\end{equation}

\noindent where $\alpha_s$ and $\delta_t$ are industry- and year-fixed effects, respectively.
The dependent variable, $\frac{NB_{s,t}}{N_{s,t}}$, is the share of newbies in industry $s$ in year $t$, where $NB_{s,t}$ is the number of newbies and $N_{s,t}$ the total number of firms in industry $s$ in year $t$. As in the previous sections, $BC^Z_{s,t-1}$, and $BN^Z_{s,t-1}$ respectively, is the share of newly granted credit from banks, and public debt markets, to industry $s$ in year $t$. The terms in the second row of Equation \eqref{equ:cleansing} each contain two further indicator variables, $bank.dep_{s,t-1}$ and $bond.dep_{s,t-1}$. In the firm-level regressions of previous sections these indicators were time-invariant. Here, we allow them to vary over time. That is, industry $s$ is classified as being \textit{bank-dependent} in year $t$ if that year's industry inflows of fresh bank-credit exceeded the amount of debt taken up via capital markets in the form of bonds and notes. Equivalently, an industry is classified as being \textit{capital market dependent} if the reverse is true. Since we again discriminate between short- and long-term debt, these indicators are conditional on the respective maturity bucket. Therefore, industry $s$ could be classified as being bank-dependent for short-term maturities, while resorting predominantly to public debt markets for longer-term funding.
Lastly, since allowing for persistence in the share of newbies does not come at the cost of losing a significant number of observations\footnote{This is not the case for the firm-level regressions, which is why we abstain from applying a similar methodology in Sections \ref{sec:empirics_prod}, \ref{sec:empirics_capgrowth}, and \ref{sec:empirics_emp}}, we deploy the \cite{ArellanoBond1991} estimator to account for potential correlation between the lagged dependent variable ($\frac{NB_{s,t-1}}{N_{s,t-1}}$) and the unobserved industry-fixed effects $\alpha_s$.

\noindent Table \ref{tab:Newbies_Z_v3_AB} documents the relation between zombie-lending and entry dynamics on the industry-level. Again, coefficients describing a statistically significant negative relationship between zombie-lending and entry-dynamics are not ample. However, the negative coefficients on $BC^Z_{s,t-1}$ in Model (1) and on $bank.dep_{s,t-1} \times BC^Z_{s,t-1}$ in Model (2) are consistent with the prevailing narrative. Industries, which have a larger share of total credit being granted to non-viable firms, experience less newbies entering the market subsequently. 
The period mismatch of one year between dependent and independent variables, allows us to establish a chain of causality by asserting that banks' participation in zombie-lending drags on a industry's entrance dynamics. 
In the case of short-term debt, the effect amounts to a -2.1 percentage point lower share of newbies in bank-dependent industries for every percentage point in the share of new zombie-to-total industry-credit that has been intermediated by banks.
Moving over to longer maturities, we see a similar story. However, with the coefficient on $BC^Z_{s,t}$ turning statistically significant, the negative spillovers do not only affect the subgroup of bank-dependent industries anymore. The results hint at a general tendency across all industries to see less firms entering their markets ($\beta_{BC} = -0.041 \sim -4.1\%$) with a rise in banks' previous period's engagement in zombie-lending relative to their total lending activities. 
%The results in Table \ref{tab:Newbies_Z_v3_AB} further allow us to quantify the multipliers of reduced entry-dynamics for each additional percentage point in the share of zombie-to-total industry-credit inflows. These amount to -2.3 for short-term lending and -3.1 in the case of long-term contracts. 

\begin{table}[h!]
    \begin{center}
    \captionsetup{justification=centering}
    \footnotesize
\caption{\label{tab:Newbies_Z_v3_AB} Regressions Results: Zombie-Lending \& Business Dynamism \\ Zombie-Definition $Z^{NAR}$}

\begin{tabular}{l ccc} \hline

 Maturity ($m$) & & 1Q  $\leq m \leq$ 4Q & 5Q  $\leq m \leq$ 40Q\\
 \cmidrule(lr){3-3} \cmidrule(lr){4-4}
 & & (1)  & (2) \\
Variables & & $\frac{NB_{s,t}}{N_{s,t}}$ & $\frac{NB_{s,t}}{N_{s,t}}$  \\ \hline \addlinespace[5pt]

\addlinespace[10pt]

%%%%%%%%%%%%%%%%%%%%%%%%%%%%%%%%%%%%%%%%%%%
$ BC^Z_{s,t-1} $ & &
0.004 &  -0.041***   \\ \addlinespace[5pt]

$ BN^Z_{s,t-1} $ & & 
0.011 &  0.049   \\ \addlinespace[5pt]
%%%%%%%%%%%%%%%%%%%%%%%%%%%%%%%%%%%%%%%%%%%

\addlinespace[10pt]

%%%%%%%%%%%%%%%%%%%%%%%%%%%%%%%%%%%%%%%%%%%
$ bank.dep_{s,t-1} \times BC^Z_{s,t-1} $ & &
-0.021** &  -0.005   \\ \addlinespace[5pt]

$ bond.dep_{s,t-1} \times BN^Z_{s,t-1} $ & & 
0.013 &  -0.042    \\ \addlinespace[5pt]
%%%%%%%%%%%%%%%%%%%%%%%%%%%%%%%%%%%%%%%%%%%

\addlinespace[10pt]

Years & & 2002-2020 & 2002-2020 \\ 
Observations & & 326 & 338  \\
Industries & & 18 & 18  \\
Fixed Effects & & X &  X   \\
Controls & & X &  X   \\

\hline
\addlinespace[1pt]
\end{tabular}
\end{center}
\noindent \scriptsize Notes: Each estimation includes industry- and year-fixed effects. Standard errors are clustered at the industry-level. 
We restrict the sample to only those companies that reported either one of the bank-credit instruments, $BL$ or $RC$, or an issuance of \textit{bonds \& notes} (BN) at least once over the period 2002 - 2019.  $ BC^{Z}_{s,t}$ and $BN^{Z}_{s,t} $ represent the amount of bank-credit, and bonds and notes respectively, of industry $s$ sitting with zombies in year $t$. $B_{s,t}$ and $CM_{s,t}$ are indicators with a value of one, if industry $s$ is classified as bank-dependent, and capital-market dependent respectively, in year $t$. These classifications may differ for different maturity buckets. Controls include a lag of the dependent variable $\frac{NB_{s,t-1}}{N_{s,t-1}}$. We use the \cite{ArellanoBond1991} estimator to account for correlation between the lagged dependent variable and unobserved industry-fixed effects. Robust standard errors in parentheses: *** p$<$0.01, ** p$<$0.05, * p$<$0.1 \\
\end{table}

\noindent Given the observed negative spillovers between zombie-lending and entry dynamics at the industry-level, the question arises about how these effects materialize on the firm-level.
One hypothesis describes the ambivalent nature of providing non-viable firms with vital liquidity \citep{AcharyaEtAl2019} and the trade-off between short-run damage-control but long-term efficiency losses \citep{AcharyaLenzuWang2021}. With non-viable firms remaining afloat, fewer workers are laid off compared to the case of liquidation. On the one hand, this implies higher aggregate employment relative to the liquidation scenario. On the other hand, labor supply is scarce, and wages abstain from dropping. %New entrants then face a shortage of labor supply and higher labor costs with non-viable establishments to attract workers at an elevated wage level.
The implications are twofold: the shortage of labor and artificially elevated industry-wages deter newbies from entering, leaving productive labor force locked up at non-viable incumbents. This hoarding of labor capacity does ultimately not only result in a within-industry misallocation of labor, but also impedes business dynamism.

\noindent The following regression shall shed some light on the assertions made above, by looking deeper into the relationship between zombie-lending and employment-growth of newbies:

\vspace{-0.5cm}
\begin{align} \label{equ:EmploymentNewbies_Zlending}
\begin{tabular}{lll}
    \(\Delta EMP_{i,t} = \) & 
    \(\beta_{BC} \; NB_{i,t-1} \times BC^Z_{s,t} \) &
    \(+ \; \beta_X \; X_{i,t-1}\)
     \\ \addlinespace[5pt] 
    & \(+ \; \beta_{BN} \; NB_{i,t-1} \times BN^Z_{s,t}\) 
    & \(+ \; \alpha_i \; + \; \delta_t \; + \; \varepsilon_{i,t}\)
\end{tabular}
\end{align}

\noindent where $\Delta EMP_{i,t}$ is the symmetric growth-rate of employment as outlined in Section \ref{sec:empirics_emp}. The set of controls, $X_{i,t-1}$, includes a lagged indicator for firm $i$ being a newbie ($NB_{i,t-1}$), as well as a firm's lagged firm size ($log\left(AT_{i,t-1}\right)$), its lagged asset tangibility ($PPENT_{i,t-1} / AT_{i,t-1}$), and sales growth. The zombie-lending variables $BC^Z_{s,t}$ and $BN^Z_{s,t}$ are defined as above.

\noindent Table \ref{tab:NewbiesEMPgr_Z_v3} shows results for short- and long-term lending, and for two different -- though related -- measures of employment growth: 
Models (1) and (2) are based on the symmetric growth rate and the latter two models deploy the first-difference of the log-level of employment.
Again, not abundant arguments for rejecting the null hypothesis of zombie-lending not having any effect on employment-growth among newbies. Yet, some statistically significant effect emerge in Models (1) and (3), which hints at negative spillovers emerging from the bond-market in the form of short-term contracts. The magnitude of the coefficient in Model (3) suggests newbies' employment-growth to decline by -1.7\%\footnote{See: $SD \times \beta_{BC} \equiv 0.046 \times \left(-0.374\right) \times 100 = -1.72\% $} for every one standard deviation increase in the share of fresh zombie-credit taken up via the bond-market.

\begin{table}[h!]
    \begin{center}
    \captionsetup{justification=centering}
    \footnotesize
\caption{\label{tab:NewbiesEMPgr_Z_v3} Regressions Results: Newbies, Employment Growth \& Zombie-Lending \\ Zombie-Definition $Z^{NAR}$}

\begin{tabular}{l ccccc} \hline
 \addlinespace[2pt]
 Maturity ($m$) & & 1Q  $\leq m \leq$ 4Q & 5Q  $\leq m \leq$ 40Q & 1Q  $\leq m \leq$ 4Q & 5Q  $\leq m \leq$ 40Q 
\\
 \cmidrule(lr){3-3} \cmidrule(lr){4-4} \cmidrule(lr){5-5} \cmidrule(lr){6-6}
 & & (1) & (2) & (3) & (4) \\
Variables & & $\Delta EMP_{i,t}$ & $\Delta EMP_{i,t}$ & $\Delta log \left(EMP_{i,t}\right)$ & $\Delta log \left(EMP_{i,t}\right)$  \\  \addlinespace[2pt] \hline \addlinespace[5pt]

%%%%%%%%%%%%%%%%%%%%%%%%%%%%%%%%%%%%%%%%%%%
$ NB_{i,t-1} $ & &
0.054*** &  0.056*** & 
0.059*** & 0.061*** \\ \addlinespace[5pt]
%%%%%%%%%%%%%%%%%%%%%%%%%%%%%%%%%%%%%%%%%%%

\addlinespace[10pt]

%%%%%%%%%%%%%%%%%%%%%%%%%%%%%%%%%%%%%%%%%%%
$ NB_{i,t-1} \times BC^Z_{s,t-1} $ & &
0.094 &  -0.594  & 
0.138 & -0.977 \\ \addlinespace[5pt]

$ NB_{i,t-1} \times BN^Z_{s,t-1} $ & & 
-0.317* &  0.143  & 
-0.374** & 0.450 \\ \addlinespace[5pt]
%%%%%%%%%%%%%%%%%%%%%%%%%%%%%%%%%%%%%%%%%%%

\addlinespace[10pt]

Years & & 2002-2020 & 2002-2020 & 2002-2020 & 2002-2020\\ 
Observations & & 68,180 & 68,273 & 68,180  & 68,273 \\ 
Firms & & 8,622 & 8,625 & 8,622 & 8,625 \\   
Fixed Effects & & X & X & X & X \\
Controls & & X &  X &  X & X \\
Within-$R^2$ & & 0.14 & 0.14 & 0.13 & 0.13 \\ 

\hline
\addlinespace[1pt]
\end{tabular}
\end{center}
\noindent \scriptsize Notes: Each estimation includes firm-, sector-, year- and sector-year-fixed effects. Standard errors are clustered at the firm-level. 
%We restrict the sample to only those companies that reported either one of the bank-credit instruments, $BL$ or $RC$, or an issuance of \textit{bonds \& notes} (BN) at least once over the period 2002 - 2019. For these variables -- and those interacted with the non-zombie-indicator ($ BC^{Z}_{i,t}, BN^{Z}_{i,t} $).
Control variables are lagged and include a proxy for firm size ($log\left(AT_{i,t-1}\right)$),  a firm's asset tangibility, i.e. the capital-to-asset ratio ($PPENT_{i,t-1} / AT_{i,t-1}$), and the first-difference of $log\left( SALE_{i,t}\right)$. Robust standard errors in parentheses: *** p$<$0.01, ** p$<$0.05, * p$<$0.1 \\ %as a proxy for business opportunities \citep{CaballeroHoshiKashyap2008,McGowanAndrewsMillot2018}. \\
\end{table}

\noindent Without posing too much weight on the single statistically significant relationship in Table \ref{tab:NewbiesEMPgr_Z_v3}, the results lend suggestive support to the hypothesis -- as also outlined in \cite{AcharyaEtAl2019} -- that zombie-prevalence deters potential candidates from entering by congesting the job market.

\section{Conclusion} \label{sec:concl}
Fueled by loose monetary policy and easy credit, the phenomenon of zombification in advanced economies has been discussed in several studies over the course of the last decade. 
The case of the U.S., however, remains largely untouched. In this study we try to fill that gap by examining the economic consequences of lending to non-viable firms. 

\noindent Our empirical analysis of publicly listed U.S. companies and their debt structure sees zombie prevalence not as a widespread phenomenon. Based on our working definition, the share of zombies barely exceeds 4\% between the early 1990s and 2020.
Nonetheless, we find zombie-lending to be of relevance in explaining the performance of non-zombies in terms of productivity, capital-, and employment growth. Negative spillovers not only emerge from banks' lending activities, but also from bond markets, as investors buy up debt securities issued by non-viable firms. Yet, the effects are most pronounced for small- and medium-sized companies. Funding characteristics, such as being bank-dependent or not having access to the bond market, amplify the negative spillovers of zombie-lending. Our analysis further shows that a differentiation among contracts with different maturities is worthwhile. While both short- and long-term lending to zombies impedes productivity of non-zombies, investment and employment-growth of non-zombies is exclusively sensitive to changes in long-term credit.
Furthermore, our results suggest that banks' zombie-lending activities congest entry- and exit-dynamics and thereby hamper the cleansing effect in the U.S. economy. 
Lastly, we find an increase in the share of new short-term industry credit in the form of bonds and notes being granted to zombies to decrease newbies' employment growth in the year after entering the market.

\noindent In a nutshell: even though prevalence of zombification and zombie-lending may not be a prominent issue among major publicly listed U.S. companies, it would be frivolous to dismiss its relevance for the broader U.S. economy. 
We find spillovers from zombie-lending to materialize predominantly among small- and medium-sized companies. 
Going forward, a more comprehensive assessment of the implications of zombification and zombie-lending, may primarily focus on the class of private and public SMEs.

\clearpage
\appendix
%\appendixpage
%\addappheadtotoc
\newcounter{saveeqn}
\setcounter{saveeqn}{\value{section}}
\renewcommand{\theequation}{\mbox{\Alph{saveeqn}.\arabic{equation}}} \setcounter{saveeqn}{1}
\setcounter{equation}{0}

\newpage
\section{Appendix}

\subsection{Computation of Total Factor Productivity} \label{sec:app_tfp}
For computing $TFP_{i,t}$, we adopt the approach in \cite{BaqaeeFarhi2020} and use the method developed in \cite{OlleyPakes1996} to estimate firm $i$'s production function.\footnote{For the computational implementation see \cite{Rovigatti2017}.} In particular, we use $log\left(SALE_{i,t}\right)$ as our outcome variable, the variable inputs $log\left(COGS_{i,t}\right)$ as the ``free'' variable, the capital stock ($log\left(PPENT_{i,t}\right)$) as the ``state'' variable. $log\left(CAPX_{i,t}\right)$ serves as the instrument for productivity.
%Figure \ref{fig:Productivity_Comparison} shows the cross-sectional means of $LP_{i,t}$ and $TFP_{i,t}$ throughout the sample period.

\subsection{Other Company Financials} \label{sec:app_variables}
This section documents the calculation of company-level variables and ratios used in the main text of the paper. Compustat identifiers are in brackets.

\begin{itemize}
    \item \makebox[3.5cm]{\textbf{Assets}:\hfill}  \textit{Total Assets} ($AT$).
    \item \makebox[3.5cm]{\textbf{Sales}:\hfill}  \textit{Net Sales} ($SALE$).
    \item \makebox[3.5cm]{\textbf{(Book) Leverage}:\hfill}  Sum of \textit{Long-Term Debt} ($DLTT$) and \\
    \makebox[3.5cm]{\hfill} \textit{Debt in Current Liabilities} ($DLC$) divided by \textit{Total Assets} ($AT$).
    \item \makebox[3.5cm]{\textbf{Asset Tangibility}:\hfill} \textit{Net Property, Plant and Equipment} ($PPENT$) divided by \\
    \makebox[3.5cm]{\hfill}  \textit{Total Assets} ($AT$).
    \item \makebox[3.5cm]{\textbf{CapX / Assets}:\hfill} \textit{Capital Expenditures} ($CAPX$) divided by \textit{Total Assets} ($AT$).
    \item \makebox[3.5cm]{\textbf{ROA}:\hfill} \textit{Net Income} ($IB$) divided by \textit{Total Assets} ($AT$).
    \item \makebox[3.5cm]{\textbf{Age}:\hfill} Difference between the year under observation and the year \\ 
    \makebox[3.5cm]{\hfill} when the company was first listed on Compustat.
    \item \makebox[3.5cm]{\textbf{Employees}:\hfill} \textit{Number of Employees} ($EMP \times 10^3$).
    \item \makebox[3.5cm]{\textbf{Capital}:\hfill} \textit{Net Property, Plant and Equipment} ($PPENT$).
    \item \makebox[3.5cm]{\textbf{R\&D Intensity}: \hfill} Expenses for Research and Development ($XRD$) divided by \\ 
    \makebox[3.5cm]{\hfill} \textit{Total Assets} ($AT$).
    \item \makebox[3.5cm]{\textbf{Cash Ratio} \hfill} Cash and Short-Term Investments (CHE) divided by \\
    \makebox[3.5cm]{\hfill} \textit{Total Liabilities}: ($LT$).
    % \item \textit{Networth}: The difference between a firm's \textit{Total Assets} ($AT$) and its \textit{Total Liabilities} ($LT$).
\end{itemize}

\noindent Not directly used in any of the estimations, but part of Tables \ref{tab:SumStat_Compustat_ZvsZ2} and \ref{tab:SumStat_Compustat_v3}, we compute \textbf{Value Added} as the difference between \textit{Net Sales} ($SALE$) and \textit{Materials}, where \textit{Materials} is the difference between \textit{Total Expenses} and \textit{Labor Costs}. \textit{Total Expenses} are computed as \textit{Net Sales} ($SALE$) - \textit{Operating Income Before Depreciation} ($OIBDP$). \textit{Labor Costs} are \textit{Staff Expense} ($XLR$). In case, $XLR$ is not reported, we proxy \textit{Labor Costs}, $LC^P$, by the product of \textit{Number of Employees} ($EMP \times 10^3$) and \textit{annual labor costs per capita in the U.S. manufacturing sector}, which itself is derived from \textit{Average Hourly Earnings of Production and Nonsupervisory Employees, Manufacturing} (FRED: $CES3000000008$) and \textit{Average Weekly Hours of Production and Nonsupervisory Employees, Manufacturing} (FRED: $AWHMAN$).

\subsection{Additional CapIQ Statistics} \label{app:capIQ_stats}

Table \ref{tab:SumStat_CapIQ_AcceptanceRates} shows the share of of first-time granted debt contracts, whose face value did not exceed the firm's reported total debt in the corresponding annual files in Compustat. A reporting was only considered eligible for the empirical analysis if its reported face value did not exceed the borrower's reported total debt.

\begin{table}[h!]
    %\begin{center}
    \centering
    \begin{threeparttable}
    \captionsetup{justification=centering}
    \footnotesize
\caption{\label{tab:SumStat_CapIQ_AcceptanceRates} Acceptance Rates of Debt Obligations by Maturities -- Full Sample: 2002-2020}

\begin{tabular}{l c cc c cc  c cc} \hline

& & \multicolumn{2}{c}{Bank/Term Loans} & & \multicolumn{2}{c}{Revolving Credit Facility} & & \multicolumn{2}{c}{Bonds and Notes}  \\ 
\cmidrule(lr){3-4} \cmidrule(lr){6-7} \cmidrule(lr){9-10}
& & Total Obs. & Accepted 
& & Total Obs. & Accepted 
& & Total Obs. & Accepted \\ 
\cmidrule(lr){3-3} \cmidrule(lr){4-4} \cmidrule(lr){6-6} \cmidrule(lr){7-7} \cmidrule(lr){9-9} \cmidrule(lr){10-10} 

%%%%%%%%%%%%%%%%%%%%%%%%%%%%%%%%%%%%%%%%%%%
1Q  $\leq m \leq$ 4Q  & &
14,174 & 76.61\% & &
17,834 & 69.71\% & &
34,302 & 80.01\% \\ \addlinespace[5pt]
%%%%%%%%%%%%%%%%%%%%%%%%%%%%%%%%%%%%%%%%%%%

%%%%%%%%%%%%%%%%%%%%%%%%%%%%%%%%%%%%%%%%%%%
5Q  $\leq m \leq$ 8Q & &
10,831 & 84.81\% & &
12,876 & 71.17\% & &
28,212 & 87.15\% \\ \addlinespace[5pt]
%%%%%%%%%%%%%%%%%%%%%%%%%%%%%%%%%%%%%%%%%%%

%%%%%%%%%%%%%%%%%%%%%%%%%%%%%%%%%%%%%%%%%%%
9Q  $\leq m \leq$ 20Q & &
30,425 & 85.26\% & &
33,630 & 70.04\% & &
75,140 & 88.94\% \\ \addlinespace[5pt]
%%%%%%%%%%%%%%%%%%%%%%%%%%%%%%%%%%%%%%%%%%%

%%%%%%%%%%%%%%%%%%%%%%%%%%%%%%%%%%%%%%%%%%%
21Q  $\leq m \leq$ 40Q & &
18,284 & 87.85\% & &
17,673 & 65.13\% & &
75,449 & 93.11\% \\ \addlinespace[5pt]
%%%%%%%%%%%%%%%%%%%%%%%%%%%%%%%%%%%%%%%%%%%

\hline

\end{tabular}
%\end{center}
\begin{tablenotes}
\scriptsize \item[] Notes: We show the fraction of newly reported debt obligations in company filings in Compustat's Capital-IQ database in the years 2002-2020, which passed the following data-cleaning procedure: an reporting of a newly reported debt obligation is only accepted to be considered in the empirical analysis if its face value does not surpass the company's total debt reported in the annual company filings.
\end{tablenotes}
\end{threeparttable}
\end{table}

\subsection{Small- \& Medium-Sized Enterprises and Subgroups} \label{app:CUS_SMEs}

See Section \ref{sec:empirics_prod} for details on the definition of SMEs and the corresponding subgroups.

\begin{figure}[h!]
    %\begin{center}
    \centering
    \caption{\footnotesize{Share of SMEs and Corresponding Subgroups}} \label{fig:CUS_SMEs}
        \begin{threeparttable}
        \includegraphics[width=0.75\textwidth, trim = 0mm 40mm 0mm 45mm, clip]{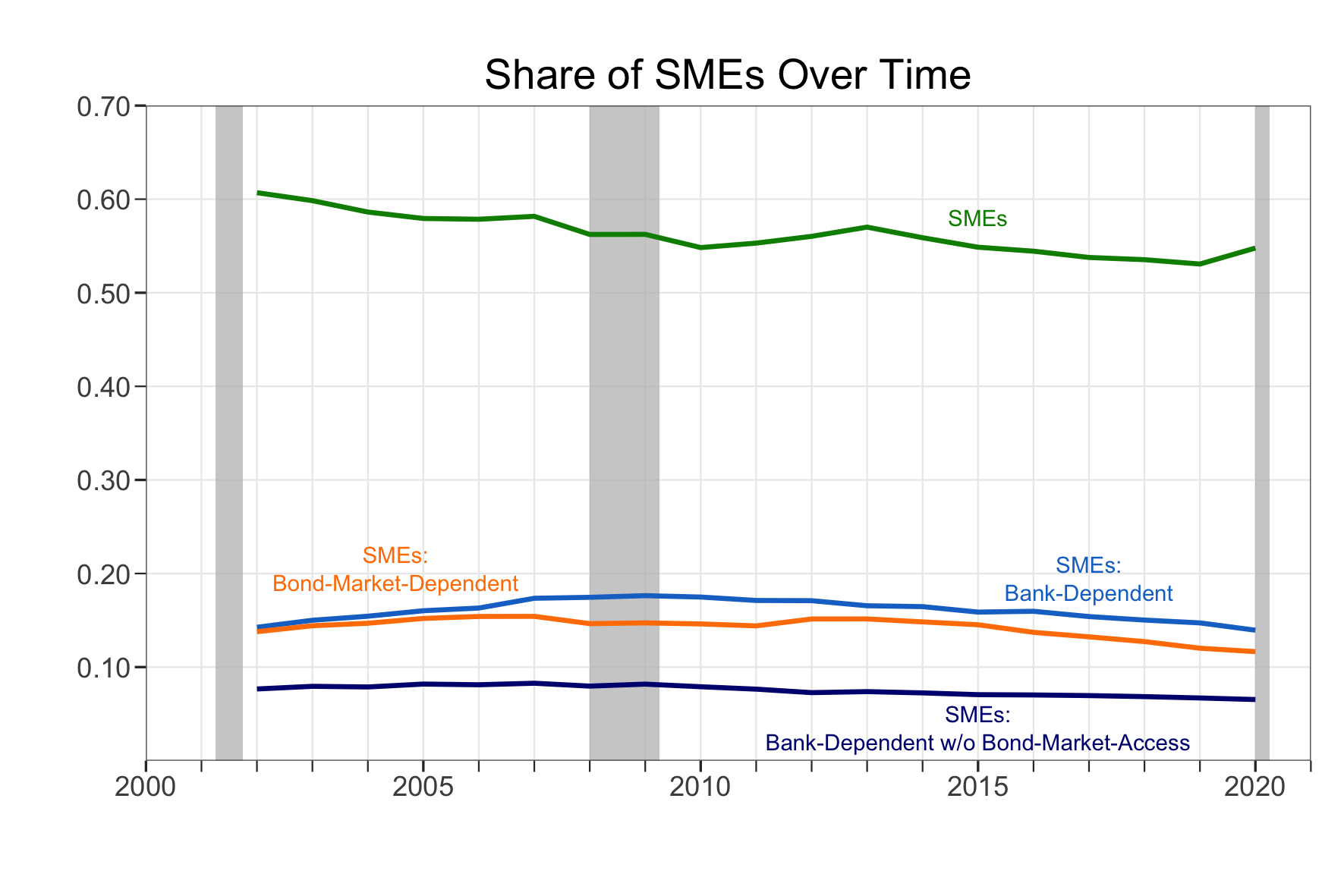}
    %\end{center}
    \vskip 5pt
    \begin{tablenotes}
    \scriptsize
\item[]  \hspace{3cm} Notes: Shaded areas mark NBER recessions. %\\ 
%\hspace{4cm} The definition of SMEs follows \cite{ChodorowReich2013}. \\ 
%\hspace{4cm} For the definition of the corresponding subgroups, see Section \ref{sec:empirics_prod}.
 \end{tablenotes}
 \end{threeparttable}
\end{figure}

\clearpage
\bibliographystyle{chicago}
\bibliography{bibliography}

\end{document}